\documentclass[floats,floatfix,showpacs,amssymb,prd,twocolumn,superscriptaddress,nofootinbib,nolongbibliography,reprint,preprintnumbers]{revtex4-1}

\usepackage{amssymb,amsmath,verbatim,mathtools,needspace,enumitem,etoolbox,graphicx,physics,microtype,afterpage,xspace,tabularx,lmodern,multirow,braket}
\usepackage{gensymb}
\usepackage[normalem]{ulem}
\usepackage{booktabs}
\usepackage[dvipsnames, usenames]{xcolor}
\definecolor{linkcolor}{rgb}{0.0,0.3,0.5}
\usepackage[unicode, colorlinks=true, linkcolor=linkcolor, citecolor=linkcolor, filecolor=linkcolor, urlcolor=linkcolor, linktocpage, breaklinks]{hyperref}
\usepackage[all]{hypcap}
\usepackage{subfigure}
\usepackage[T1]{fontenc}
\usepackage[utf8]{inputenc}
\usepackage[usenames,dvipsnames]{xcolor}
\hypersetup{colorlinks=true,citecolor=romared,linkcolor=romared,urlcolor=romared}

\definecolor{romared}{RGB}{142,0,28}

\newcommand{\be}{\begin{equation}}
\newcommand{\ee}{\end{equation}}

\def\be{\begin{equation}}
\def\ee{\end{equation}}
\newcommand{\beq}{\begin{eqnarray}}
\newcommand{\eeq}{\end{eqnarray}}

\usepackage{aas_macros}
\usepackage{makecell}
\usepackage{soul}
\usepackage{amssymb}
\usepackage{longtable}
\usepackage{lipsum}

\newcolumntype{Y}{>{\centering\arraybackslash}X}

\begin{document}

\title{Theoretical Aspects of Direct Waves in Kerr Black Holes:\\ Pole-Splitting Method for Ringdown Analysis}

\author{Nao Nakamoto} 
\affiliation{Department of Physics, Kyoto University, Kyoto 606-8502, Japan}

\author{Naritaka Oshita}
\affiliation{Department of Physics, Kindai University, Osaka 577-8502, Japan}
\affiliation{RIKEN iTHEMS, Wako, Saitama, 351-0198, Japan}

\author{Hiroki Takeda}
\affiliation{The Hakubi Center for Advanced Research, Kyoto University, Kyoto 606-8501, Japan}
\affiliation{Department of Physics, Kyoto University, Kyoto 606-8502, Japan}

\preprint{RIKEN-iTHEMS-Report-26}

\begin{abstract}
We develop a theoretical description of direct waves (DWs) from a particle plunging into a Kerr black hole in the extreme-mass-ratio regime.
A DW is characterized by a source-driven complex frequency $\omega_{\rm G}$ that reflects the particle's orbital motion, including motion inside the ergoregion.
Hence, it is expected to be a probe of rich structures of Kerr black holes.
We show that the contribution of the quasinormal-mode (QNM) pole sector to the DW is part of dynamical QNM excitation.
We also show that the apparent horizon-mode poles cancel at every order of a consistent near-horizon expansion of the trajectory, homogeneous radial solution, and complete point-particle source.
This result assumes that the point particle follows a geodesic near the horizon and leaves open the possibility of a transient DW while the particle passes near the ergosurface or the vicinity of the potential barrier.
To investigate this possibility in simulated plunge waveforms, we develop a \emph{pole-splitting} method that decomposes the transfer function into QNM-pole and non-pole parts.
The resulting waveform decomposition avoids the frequency-dependent time shift (group delay) introduced by the QNM filtering.
For quasi-circular plunges into moderately and rapidly rotating Kerr black holes, the frequency and decay rate of the non-pole waveform show qualitative agreement with our theoretical prediction over finite time intervals.
These results support the potential use of DWs as probes of the ergoregion and gravitational redshift effect around black holes.
\end{abstract}

\maketitle

%%%%%%%%%%%%%%%%%%%%%%%%%%%%
\section{Introduction}
%%%%%%%%%%%%%%%%%%%%%%%%%%%%

Gravitational waves (GWs) from binary black-hole (BBH) coalescences probe dynamical spacetime in the strong-field regime~\cite{LIGOScientific:2026wfs}. 
The remnant of a BBH merger settles to a stationary Kerr black hole, whose late-time radiation is modeled by a superposition of quasinormal modes (QNMs), damped oscillations with
complex frequencies determined by the remnant mass and spin~\cite{Vishveshwara:1970zz,Chandrasekhar:1975zza,Berti:2009kk}. 
Measuring several modes therefore tests the Kerr geometry, i.e., black-hole spectroscopy~\cite{Dreyer:2003bv,Berti:2005ys,LIGOScientific:2026wpt,LIGOScientific:2025wao}.

A superposition of QNMs provides a successful leading description of black-hole ringdown, but the full response can also contain non-QNM contributions. 
In Leaver's contour-integral representation of the retarded Green's function~\cite{Leaver:1986gd}, the response comprises residues at the QNM poles, a branch-cut integral giving the late-time tail, and a large-arc contribution giving the prompt response. 
Moreover, the excitation coefficients of the QNMs can themselves be dynamical, depending on the source trajectory~\cite{DeAmicis:2025xuh, Weller:2026wjh}.

Recent works have focused on a source-driven signal in the plunge stage in the extreme-mass-ratio regime, referred to as the \emph{direct wave} (DW). 
Building on earlier studies of the horizon mode and its screening for rotating black holes~\cite{Mino:2008at,Zimmerman:2011dx}, Oshita {\it et al.}~\cite{Oshita:2025qmn}
proposed that a signal sourced around the ergoregion or the light ring contains a characteristic component modulated by the plunging motion.
In this description, the frequency evolution carries the effects of frame dragging and gravitational redshift, so that the horizon angular velocity $\Omega_{\rm H}$ and the surface gravity $\kappa$ may be imprinted on the DW, although the selective screening against the horizon mode makes it decay rapidly~\cite{Zimmerman:2011dx,Oshita:2025qmn, Kubota:2026qjh}.

Since then, the DW has been studied from several directions. 
Analytic work has examined whether such a component survives in the observable waveform, and how the linear response should be decomposed in the first place~\cite{Kuntz:2026xep,Ma:2026hcb,Su:2026fvj,Su:2026gmp,Han:2026cgd}.
Analyses of gravitational-wave data, based on the removal of the dominant QNMs by rational filters~\cite{Ma:2022wpv,Lu:2025mwp}, have reported a DW component in GW250114 and used it to infer black-hole quantities~\cite{Lu:2025vol,Chung:2026eph,LIGOScientific:2025rid}. 
Numerical-relativity waveforms have been used to extract the component from comparable-mass waveforms and to test its relation to $\Omega_{\rm H}$ and $\kappa$~\cite{Kankani:2026byb,Kankani:2026kst,Dyer:2026yex,Sun:2026mto,Cheung:2026gfd}, although it is still non-trivial whether the naive extension of the picture of DW can be validated. 
Some of these works agree that a time-dependent non-QNM component can be extracted, while its relation to black-hole quantities and the appropriate modeling of DWs remain under debate.

Recently, Kubota {\it et al.}~\cite{Kubota:2026qjh} demonstrated the cancellation of source-induced horizon-mode poles for a particle plunging into a Kerr black hole.
Their calculation expands the orbital phase to all orders in the distance from the horizon, while retaining only the leading-order near-horizon expressions for the homogeneous radial Teukolsky solution $R^{\rm in}$ and the source amplitudes $A_{ab}$.
If the horizon modes in the near-horizon regime completely vanish, it excludes a surviving late-time horizon mode, although it relies on an idealized geodesic point-particle description arbitrarily close to the horizon.
Higher-order terms in the radial solution can themselves have poles at the same frequencies~\cite{Motohashi:2026mbn}.
Also, the full forms of the source coefficients $A_{ab}$ are important as they may lead to the non-trivial screening at higher order of $\epsilon$.
We here consistently expand all terms in the spectrum, i.e., the phase factor evolving with the motion of the plunging particle, the in-mode homogeneous Teukolsky solution $R^{\rm in}$, and the source coefficients $A_{ab}$, and then show that the horizon-mode cancellation occurs at all orders.
Under the assumptions stated in Sec.~\ref{sec:NH-cancellation}, the horizon-mode poles therefore remain absent at every order of the near-horizon expansion.
However, it still leaves open the possibility of a transient DW while the particle traverses the ergoregion or the vicinity of the potential barrier.

To improve DW modeling and better understand their theoretical aspects, we formulate DWs in the Kerr spacetime for extreme-mass-ratio mergers.
Since the validity of the linear picture in comparable-mass mergers requires a more careful treatment, we restrict our analysis to the extreme-mass-ratio regime.
By expanding the trajectory about successive points and deforming the contour of the retarded Green's function integral for each segment, Oshita {\it et al.}~\cite{Oshita:2025qmn} derived the DW and the QNMs simultaneously as pole contributions (see Supplemental Material in Ref.~\cite{Oshita:2025qmn}). 
We obtain the complex DW frequency $\omega_{\rm G}(u)$ in this way.
We also extend our analysis and discuss its relation to the dynamically excited QNMs~\cite{DeAmicis:2025xuh}. 
We use this construction to examine how the plunging motion is reflected in the waveform over finite time intervals.

To verify our analysis in the linear-perturbation regime, we analyze the GW signal sourced by a particle plunging into a Kerr black hole and extract its DW component.
We do not rely on the QNM filtering, which may cause a frequency-dependent time shift, i.e., a group delay.
We instead introduce a spectral \emph{pole-splitting} procedure.
It decomposes the transfer function into a QNM-pole sector and a non-pole sector, giving two waveforms $\Psi_{\rm pole}$ and $\Psi_{\rm non}$.
\begin{figure}[t]
\centering
\includegraphics[width=1\linewidth]{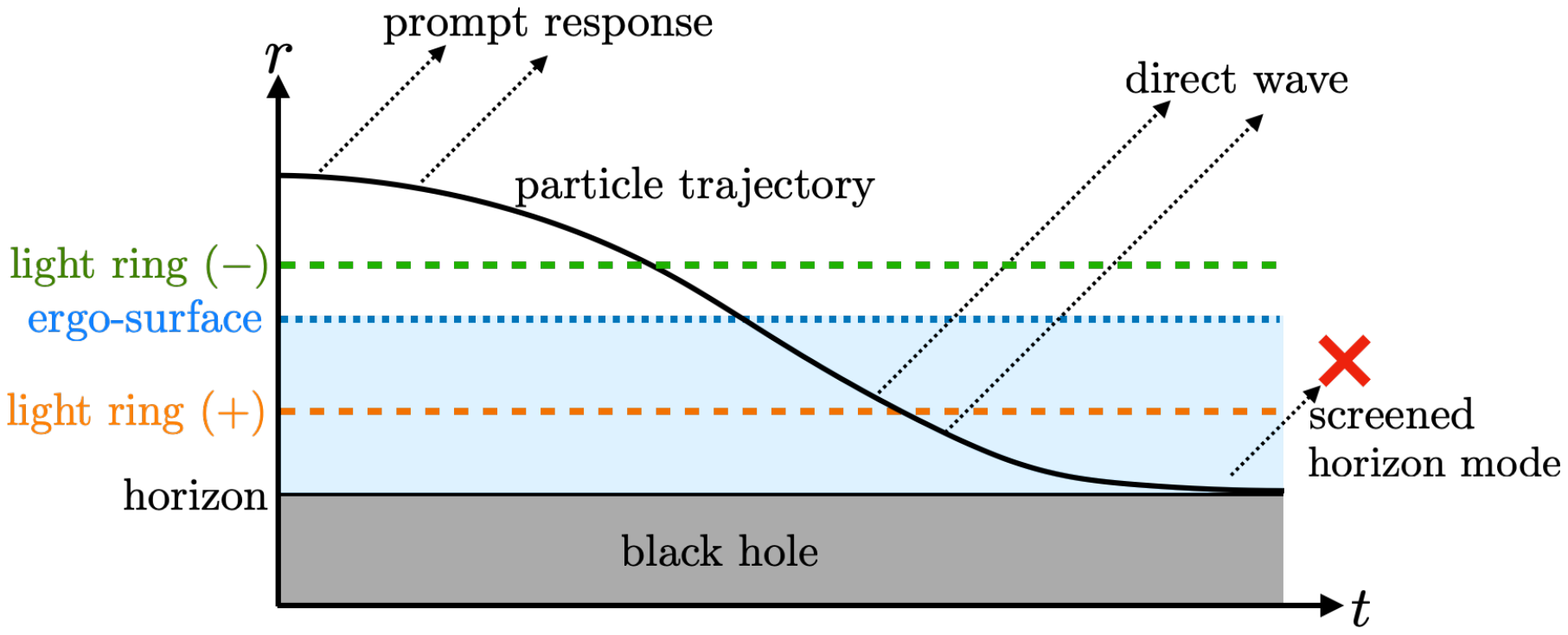}
\caption{
Schematic picture of prompt response, DW, and screened horizon mode.
}
\label{fig_DW_schematic}
\end{figure}
The pole sector contains the QNMs and may also contain part of the DW, whereas $\Psi_{\rm non}$ contains no QNMs by construction.
We therefore test whether the DW model describes the frequency evolution of the non-pole waveform by comparing it with the predicted frequency derived from the plunging particle's motion.
For quasi-circular plunges with $\chi=0.8$, $0.9$, and $0.95$ for the dominant angular mode of $\ell=m=2$, we find that $\Psi_{\rm non}$ follows $\omega_{\rm G}$, or its screened counterpart $\omega_{\rm screen}$ once the source lies inside the potential barrier.

The paper is organized as follows. 
Section~\ref{sec_Theory} reviews the near-horizon descriptions of DWs in the Teukolsky formalism and derives the source-driven frequency $\omega_{\rm G}$ following Oshita {\it et al.}~\cite{Oshita:2025qmn}, and discusses their relation to dynamically excited QNMs.
It also establishes horizon-mode pole cancellation with all orders of the near-horizon radial, source, and trajectory expansions retained under the stated assumptions.
Sec.~\ref{sec:Pole-Splitting-Method} introduces the pole-splitting method, which is a way to reveal the non-QNM component in the waveform, an alternative to the QNM filtering.
Sec.~\ref{sec:Numerical-direct-wave} applies it to plunges in Kerr, and Sec.~\ref{sec:Conclusion} is devoted to our conclusion.
In Appendix~\ref{app_detail_PSmethod}, the details of our pole-splitting method are provided.
In Appendix~\ref{app_screening_SN}, to be comprehensive, we provide the details of the screening factor both in the Teukolsky formalism and the Sasaki-Nakamura formalism.
We discuss the zeros at the horizon modes, $\omega_{\rm H}^{(n)} \coloneqq m\Omega_{\rm H} -in \kappa$ with $n=1,2,3,...$, in the transmissivity of the Kerr geometry, $1/B^{\rm in}_{\rm SN}$, in the Sasaki-Nakamura formalism for the spin field of $s=-2$.
We use $G=c=1$ throughout the manuscript.

%%%%%%%%%%%%%%%%%%%%%%%%%%%%
\section{Theory of direct wave emission}
\label{sec_Theory}
%%%%%%%%%%%%%%%%%%%%%%%%%%%%

In this section, we formulate the theoretical aspects of DW sourced by a particle plunging into a Kerr black hole.
The Kerr geometry is described in Boyer-Lindquist coordinates
$x^\mu=(t,r,\theta,\phi)$ by
\begin{align}
    ds^2 ={}& -\frac{\Delta}{\Sigma}
    \left(dt-a\sin^2\theta\,d\phi\right)^2
    +\frac{\Sigma}{\Delta}dr^2 \notag\\
    &+\Sigma d\theta^2
    +\frac{\sin^2\theta}{\Sigma}
    \left[(r^2+a^2)d\phi-a\,dt\right]^2\,,
    \label{eq:kerr-metric}
\end{align}
where
\begin{equation}
    \Delta := r^2-2Mr+a^2\,,
    \qquad
    \Sigma := r^2+a^2\cos^2\theta\,.
    \label{eq:kerr-functions}
\end{equation}
The black-hole mass and angular momentum are denoted by $M$ and $J=aM$, respectively, and the dimensionless spin is $\chi \coloneqq a/M$.
The outer event horizon is located at $r_+ \coloneqq M+\sqrt{M^2-a^2}$.

We aim to organize the theory of DW emission in this geometry, discuss its relation to dynamically excited QNMs, and examine how the DW frequency evolution can be identified in the waveform.
Sections~\ref{sec:DW_in_Kerr} and \ref{sec:source} review the role of Kerr geometry and the near-horizon source spectrum, establishing the physical setting for the subsequent analysis.
We then evaluate the response to a finite source segment using two Green's function methods.
The frequency-domain Green's function method in Sec.~\ref{subsec:dw_from_spectral} derives the DW by evaluating the source-time integral before the frequency integral.
The time-domain Green's function method in Sec.~\ref{sec:DW-analytic} reverses this order and expresses the QNM response in terms of time-dependent amplitudes.
In Sec.~\ref{subsec:DW-composition}, we apply a pole/non-pole decomposition to the frequency-domain result and show that the pole-sector contribution to the DW is part of dynamically excited QNMs.
We also explain why the non-pole sector is expected to reveal the evolution of the source-driven frequency $\omega_{\rm G}$ when the remaining integral contributions can be neglected.
This analysis provides the basis for the waveform decomposition and numerical comparisons in the following sections.
Finally, Sec.~\ref{sec:NH-cancellation} examines the cancellation of the horizon modes by expanding the source and Green's function to arbitrarily high orders in the distance from the horizon under the stated assumptions, and discusses its implications for the DW model.

%%%%%%%%%%%%%%%%%%%%%%%%%%%%
\subsection{DWs and Kerr black hole geometry}
\label{sec:DW_in_Kerr}
%%%%%%%%%%%%%%%%%%%%%%%%%%%%

Throughout this paper, we use the term DW for a source-driven component in a GW, whose frequency evolution reflects frame dragging and gravitational redshift of a Kerr black hole, while its propagation to the observer is screened by the curvature barrier associated with the light ring (FIG.~\ref{fig_DW_schematic}). 

\begin{figure}[t]
\centering
\includegraphics[width=0.95\linewidth]{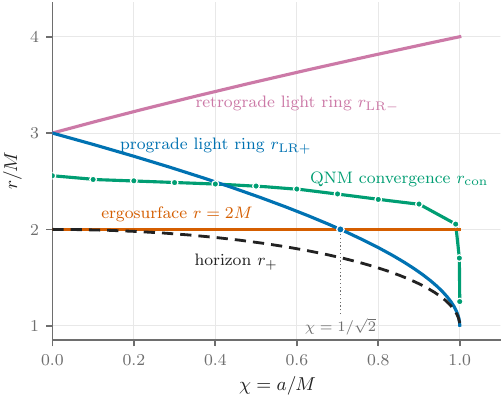}
\caption{
Spin dependence of the Kerr geometry. The radial hierarchy between the ergo-surface and the prograde light ring switches at $\chi = 1/\sqrt{2} \simeq 0.7$. 
As a reference, the QNM convergence radii for different spins, obtained in Ref.~\cite{Oshita:2026vxh}, are also shown.
}
\label{fig_Kerr_geo}
\end{figure}

Such a signal is expected to be most clearly manifested when the source passes through a region that lies inside the ergoregion and near or inside the prograde light ring, where the redshift is already substantial.
Such a situation can be realized when the black hole has a rapid spin.
As a dimensionless measure of proximity to the near-horizon regime, we use
\begin{equation}
    f(\chi, r) \coloneqq \frac{\Delta}{r^{2}+a^{2}} \propto r-r_+\,,
    \label{eq:redshift-factor}
\end{equation}
which vanishes at the outer horizon. 
In the ergoregion, the source is forced to co-rotate with the black hole; as it approaches the horizon, its angular velocity tends to $\Omega_{\rm H}$ and the real part of the emitted frequency approaches $m\Omega_{\rm H}$.

\begin{figure}[t]
\centering
\includegraphics[width=0.95\linewidth]{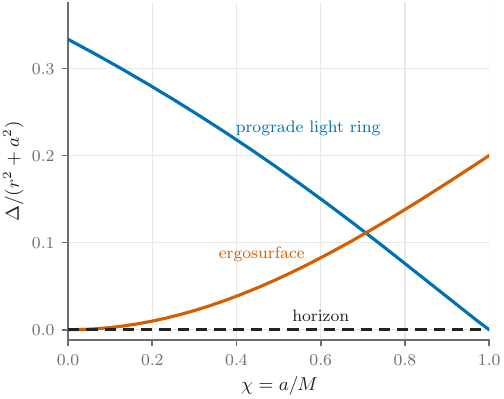}
\caption{
Redshift factor $f(\chi, r)$, defined in Eq. \eqref{eq:redshift-factor}, at the prograde light ring (blue), ergosurface (orange) and outer horizon (black dashed).
}
\label{fig_red_shift}
\end{figure}

Whether the light-ring region also lies in this strongly dragged and redshifted domain depends on the spin. 
As shown in FIG.~\ref{fig_Kerr_geo}, the prograde light ring crosses the equatorial ergosurface at $\chi=1/\sqrt{2}\simeq0.707$ and approaches the outer horizon as $\chi$ increases. 
FIG.~\ref{fig_red_shift} shows that $f$ at the prograde light ring decreases from $1/3$ and vanishes in the extremal limit. 
We see that above $\chi \simeq 0.7$, the ergoregion, the light-ring region, and the strongly red-shifted region increasingly overlap, where the redshift factor $f$ is suppressed.

We adopt a hybrid approximation in which the radial function \(R^{\rm in}_{\ell m\omega}\) and the source coefficients \(A_{\bar m\bar m j}\) are evaluated at leading order in the near-horizon expansion, while the orbital phase is Taylor-expanded around a reference time along the trajectory.

Since an analytic treatment of the Green's function around the light ring is generally difficult, we assume that the near-horizon form of the in-mode Teukolsky homogeneous solution $R^{\rm in}_{\ell m\omega}$, characterized by the factor $e^{-i(\omega-m\Omega_{\rm H})x}$, remains a useful approximation in this region, where $\Omega_{\rm H}$ is the horizon angular velocity and $x$ is the tortoise coordinate. 
This assumption may be particularly reasonable for rapidly rotating black holes, as the equatorial prograde light ring lies inside the ergoregion for $\chi>1/\sqrt{2}$ and approaches the horizon in the extremal limit. 
The leading-order near-horizon source coefficients are adopted to incorporate the screening of horizon modes discussed by Zimmerman and Chen \cite{Zimmerman:2011dx}.

In contrast, the Taylor expansion of the orbital phase is intended to capture the orbital dynamics beyond the near-horizon regime, including the region outside the light ring. 
This hybrid prescription provides a simple phenomenological model of orbit-driven waveforms.
The validity of this phenomenological prescription will be assessed in Sec.~\ref{sec:Numerical-direct-wave} through a comparison with numerical simulation waveforms.

It should be emphasized that the radius at which the screening occurs is not obvious.
On the equatorial plane, the prograde and retrograde light rings sit at different
radii.
Moreover, the effective excitation radius of the QNMs, obtained in
Ref.~\cite{Oshita:2026vxh} from the Stokes geometry and from the convergence of the excitation factors, generally differs from either light ring, and for medium
and rapid spins it lies between the two (see FIG.~\ref{fig_Kerr_geo}).
The origin of the screening therefore requires a similarly careful treatment.

Identifying a DW in an extreme-mass-ratio merger could complement QNM spectroscopy by probing the source motion and wave propagation in this strong-field region. 
With this motivation, the rest of this section develops the near-horizon description of the DW for black-hole spins $\chi > 0.7$.

%%%%%%%%%%%%%%%%%%%%%%%%%%%%
\subsection{Near-horizon trajectory and the spectrum}
\label{sec:source}
%%%%%%%%%%%%%%%%%%%%%%%%%%%%
In this subsection, we review the near-horizon source spectrum and express it in a form suitable for comparing the two Green's function methods.
The resulting expression identifies the complex frequency associated with the particle motion and separates the source dependence from the transfer function.
We define $\epsilon\coloneqq r-r_+$, which is the expansion parameter to perform the near-horizon approximation. 
We denote the particle mass by $\mu$.
The constants $E$ and $L_z$ are the energy and axial angular momentum per unit particle mass.
The source term is determined by the geodesics of a particle that is governed by the following equations in the near-horizon regime \cite{Mino:2008at}:
\begin{align}
    \frac{dr}{d\lambda} = -2Mr_+ (E-\Omega_{\rm H} L_z) + {\cal O} (\epsilon)\,,\\
    \frac{d\theta}{d\lambda} = \pm \sqrt{\Theta_0} + {\cal O}(\theta - \theta_0)\,,\\
    \frac{d\phi}{d\lambda} = \frac{a}{2\kappa} (E-\Omega_{\rm H} L_z) \frac{1}{r-r_+} + {\cal O} (\epsilon^0)\,,\\
    \frac{dt}{d\lambda} = \frac{M}{\kappa} (E-\Omega_{\rm H} L_z) \frac{r_+}{r-r_+} + {\cal O} (\epsilon^0)\,,
\end{align}
where $\lambda$ is the Mino time, defined by $d\tau = \Sigma_{\rm BL} d\lambda$ with $\Sigma_{\rm BL} \coloneqq r^2+a^2\cos^2\theta$, and $\tau$ denotes the proper time of a particle.
$\Theta_0$ is the polar potential evaluated at the angle $\theta_0$ at which the particle plunges into the horizon.
We write $\Omega_{\rm H} \coloneqq a/(2M r_+)$ for the horizon angular velocity and $\kappa \coloneqq (r_+ - r_-)/(4Mr_+)$ for the surface gravity.

Those geodesic equations admit the universal motion of the particle in the radial and azimuthal directions in the near-horizon regime:
\begin{align}
    r &= r_+ \left( 1 + e^{-2\kappa (t-t_0)} \right) + {\cal O}(\epsilon^2)\,,\label{eq:NH-r-of-t}\\
    \theta &= \theta_0 \mp \frac{\sqrt{\Theta_0} e^{- 2\kappa (t-t_0)}}{2M (E-\Omega_{\rm H} L_z)} + {\cal O}(\epsilon^2)\,,\\
    \phi &= \Omega_{\rm H} \left( t-t_0 \right) + \phi_0 + {\cal O}(\epsilon)\,,\label{eq:NH-phi-of-t}
\end{align}
where $t_0$ and $\phi_0$ are integration constants fixed by the initial condition.
As the particle near the horizon has such a universal trajectory, it also sources a GW signal with a universal feature in the near-horizon limit, i.e., the horizon mode \cite{Mino:2008at}.

We compute the radiation from the plunging particle in the Teukolsky formalism.
The scalar $\Psi_{4}$ can be decomposed as
\begin{equation}
\label{eq:psi4-decomp}
\Psi_{4} = \frac{1}{(r-ia \cos \theta)^4} \int \frac{d \omega}{2\pi} \sum_{\ell m} e^{-i \omega t + i m \phi} R_{\ell m \omega} (r) S_{\ell m \omega} (\theta)\,,
\end{equation}
where $R_{\ell m \omega} (r)$ is the radial Teukolsky variable and $S_{\ell m \omega} (\theta)$ is a spin-weighted spheroidal harmonic.
The harmonics depend on the frequency through $a\omega$.
\footnote{We follow the bilinear normalization of the spheroidal harmonics~\cite{Su:2026gmp},
\begin{equation}
\label{eq:bilinear}
    \int_0^{\pi}\bigl[S_{\ell m\omega}(\theta)\bigr]^2\sin\theta\,d\theta =1\,.
\end{equation}
In the convention of Eq.~\eqref{eq:bilinear}, we can analytically continue the real frequencies to complex frequencies.
This is the condition required by the contour deformations in Sec.~\ref{subsec:dw_from_spectral} and \ref{sec:DW-analytic}.
}

The variable $R_{\ell m \omega} (r)$ is obtained as the inhomogeneous solution to the Teukolsky equation.
The solution takes the following form in the limit of $r \to \infty$:
\begin{align}
\begin{split}
R_{\ell m \omega} (r \to \infty) &\to \frac{r^3 e^{i \omega x}}{2i \omega B_{\ell m \omega}^{\rm in}} \int_{r_+}^{\infty} dr R_{\ell m \omega}^{\rm in} \Delta^{-2} T_{\ell m \omega}\,,\\
&\eqqcolon r^3 e^{i \omega x} Z_{\ell m \omega}\,,
\end{split}
\label{spectrum_original_for_far}
\end{align}
where $T_{\ell m \omega}$ is the source term associated with the particle plunging into the Kerr black hole, and $R_{\ell m \omega}^{\rm in}$ is the homogeneous solution to the Teukolsky equation satisfying the incoming boundary condition at the horizon:
\begin{align}
R_{\ell m \omega}^{\rm in} \to 
\begin{cases}
    \Delta^2 e^{-i k_{\rm H} x} \ \text{for} \ r \to r_+\,,\\
    B_{\ell m \omega}^{\rm out} r^3 e^{i \omega x} + B_{\ell m \omega}^{\rm in} r^{-1} e^{-i \omega x} \ \text{for} \ r \to \infty\,,
\end{cases}
\label{eq:Rin_asympt}
\end{align}
where $k_{\rm H} \coloneqq \omega - m \Omega_{\rm H}$ and the tortoise coordinate $x$ is defined as
\begin{equation}
    x = r + \frac{2M r_+}{r_+ - r_-} \ln \left( \frac{r-r_+}{r_+-r_-} \right)
    + \frac{2M r_-}{r_- - r_+} \ln \left( \frac{r-r_-}{r_+-r_-} \right)\,.
\end{equation}

The leading near-horizon source spectrum contains a frequency-dependent factor that vanishes at two complex horizon frequencies.
To exhibit this factor, we write the spectrum in terms of the near-horizon source coefficients:
\begin{align}
\begin{split}
    Z_{\ell m \omega} &\simeq \frac{\mu}{2i \omega B_{\ell m \omega}^{\rm in}} \int dt e^{i \omega t - im \phi(t)}\\
    &\times \left\{ R_{\ell m \omega}^{\rm in} A_{\bar{m} \bar{m} 0} -(\partial_r R_{\ell m \omega}^{\rm in}) A_{\bar{m} \bar{m} 1} \right.\\
    &\left. ~~~~~~~~~~~~~~~~~~~~+( \partial_r^2 R_{\ell m \omega}^{\rm in}) A_{\bar{m} \bar{m} 2} \right\}\,.
\end{split}
\label{spectra_mm_form}
\end{align}
The leading expressions for $A_{\bar{m} \bar{m} 0}$, $A_{\bar{m} \bar{m} 1}$ and $A_{\bar{m} \bar{m} 2}$ follow from the point-particle source \cite{Mino:2008at, Zimmerman:2011dx}:
\begin{align}
\begin{split}
A_{\bar{m} \bar{m} 0} &= - \frac{{\cal A}S_{\ell m \omega} (\theta_0)}{r-r_+}
\left\{-i \frac{k_{\rm H}}{2\kappa} + \left( \frac{k_{\rm H}}{2\kappa} \right)^2 \right\} + {\cal O}(\epsilon^0) \,,\\
A_{\bar{m} \bar{m} 1} &= i {\cal A}S_{\ell m \omega} (\theta_0)
\frac{k_{\rm H}}{\kappa} + {\cal O}(\epsilon) \,,\\
A_{\bar{m} \bar{m} 2} &= {\cal A}S_{\ell m \omega} (\theta_0) \epsilon + {\cal O}(\epsilon^2)\,,
\end{split}
\label{mm_components}
\end{align}
with
\begin{equation}
{\cal A} \coloneqq \frac{\kappa a^2 E_{\rm ISCO}}{2 \sqrt{2 \pi} M r_+} \frac{r_+ - ia \cos \theta_0}{r_+ + ia \cos \theta_0} \sin^2 \theta_0 \,.
\end{equation}
Substituting Eq.~\eqref{mm_components} into Eq.~\eqref{spectra_mm_form}, we obtain
\begin{align}
\begin{split}
Z_{\ell m \omega} &= \frac{\mu \tilde{\cal A}S_{\ell m \omega} (\theta_0)}{2i\omega B_{\ell m \omega}^{\rm in}}\left(k_{\rm H} + i \kappa \right) \left(k_{\rm H} +2 i \kappa \right)\\
&\times \int dt e^{i \omega t - i m \phi(t)} \frac{\epsilon}{r_+} e^{-i k_{\rm H} x(t)} + {\cal O}(\epsilon^2)\,,
\end{split}
\label{spectrum_final_form}
\end{align}
where 
\begin{equation}
\tilde{\cal A} \coloneqq - \frac{8Mr_+^2 \kappa a^2 E_{\rm ISCO}}{\sqrt{2 \pi}} \frac{r_+ - ia \cos \theta_0}{r_+ + ia \cos \theta_0} \sin^2 \theta_0\,.
\end{equation}
The factor $\left(k_{\rm H} + i \kappa \right) \left(k_{\rm H} +2 i \kappa \right)$ in Eq.~\eqref{spectrum_final_form} vanishes at the two horizon frequencies $m\Omega_{\rm H}-i\kappa$ and $m\Omega_{\rm H}-2i\kappa$.
We use screening to denote the suppression or cancellation of a source-frequency contribution by a zero of a source or response factor.
Section~\ref{sec:NH-cancellation} examines horizon-mode pole cancellation with all near-horizon orders retained.

To describe the frequency evolution along the plunge, we retain the orbital functions $\phi(t)$ and $x(t)$ in Eq.~\eqref{spectrum_final_form} and expand them over short time intervals, as described in Sec.~\ref{sec:DW_in_Kerr}.
Following Ref.~\cite{Oshita:2025qmn}, we divide the integral in Eq.~\eqref{spectrum_final_form} into pieces $[t_i, t_{i+1}]$ and decompose the spectrum $Z_{\ell m \omega}$ as $Z_{\ell m \omega} = \sum_i z_{\ell m \omega [i]}$ with
\begin{align}
\begin{split}
z_{\ell m \omega [i]} &\simeq \frac{\mu \tilde{\cal A}S_{\ell m \omega}(\theta_0)}{2i\omega B_{\ell m \omega}^{\rm in}}\left(k_{\rm H} + i \kappa \right) \left(k_{\rm H} +2 i \kappa \right) \\
&\times \int_{t_i}^{t_{i+1}} d t \frac{\epsilon}{r_+}
\exp \left[{i \omega t - i m \phi(t)}  -i k_{\rm H} x(t)\right] \,.
\end{split}
\end{align}
Let us perform the Taylor expansion for the particle trajectory around $t=t_i$:
\begin{align}
\phi (t) &= \phi (t_i) + \Omega_i \Delta t + {\cal O}[\Delta t^2]\,,\\
x (t) &= x (t_i) + \beta_i \Delta t + {\cal O}[\Delta t^2] \,,
\end{align}
where $\Omega \coloneqq d \phi / dt$, $\beta \coloneqq dx / dt$, and $\Delta t \coloneqq t-t_i$.
The lower index $i$ stands for a quantity at $t = t_i$.
We retain terms linear in $\Delta t$, with $\Omega_i$ and $\beta_i$ evaluated along the orbit.
Each interval must be short enough that the neglected terms in the exponent remain small over the relevant frequencies.

At leading order near the horizon, the tortoise coordinate gives
\begin{equation}
\epsilon / r_+ = 4M \kappa e^{-2 r_+ \kappa} e^{2 \kappa x}\,.
\end{equation}
We then have
\begin{widetext}
\begin{align}
\begin{split}
z_{\ell m \omega [i]} &\simeq \frac{\mu \tilde{\cal A}S_{\ell m \omega}(\theta_0)} {2i\omega B_{\ell m \omega}^{\rm in}}\left(k_{\rm H} + i \kappa \right) \left(k_{\rm H} +2 i \kappa \right) 4M \kappa e^{-2r_+ \kappa} \\
& \times \exp \left[{i \omega t_i - i m \phi(t_i)}  -i \left(k_{\rm H} +2i \kappa \right) x(t_i)\right] \times \int_{t_i}^{t_{i+1}} d t
\exp \left[{i \omega \Delta t - i m \Omega_i \Delta t}  -i \left(k_{\rm H} + 2 i \kappa \right) \beta_i \Delta t \right]\,,
\end{split}
\end{align}
\end{widetext}
We distinguish the retarded-time coordinate along the source trajectory, $u_{\rm s}(t)\coloneqq t-x(t)$, from the observer's retarded time $u$.
For the $i$-th source segment, we define $u_i\coloneqq u_{\rm s}(t_i)$ and use $\Delta u\coloneqq u_{\rm s}(t)-u_i$ as the integration variable.
The linear trajectory approximation gives $\Delta u\simeq(1-\beta_i)\Delta t$.
The segment duration is $\Delta u_i\coloneqq u_{\rm s}(t_{i+1})-u_{\rm s}(t_i)=u_{i+1}-u_i$.
This gives
\begin{align}
\begin{split}
\label{eq:spectrum_fandamental}
z_{\ell m \omega [i]} &\simeq \frac{\mu \tilde{\cal A}S_{\ell m \omega}(\theta_0)} {2i\omega B_{\ell m \omega}^{\rm in}}\left(k_{\rm H} + i \kappa \right) \left(k_{\rm H} +2 i \kappa \right) \frac{4M \kappa e^{-2r_+ \kappa}}{1-\beta_i} \\
& \times \exp \left[i \omega u_i -i \int^{u_i}  \omega_{\rm G} du \right] \\
&\times \int_{0}^{\Delta u_i} d \Delta u
\exp \left[
i (\omega - \omega_{{\rm G},i}) \Delta u \right]\,,
\end{split}
\end{align}
where 
\begin{equation}
    \omega_{\rm G} \coloneqq m \frac{\Omega - \beta \Omega_{\rm H}}{1- \beta} + 2i \kappa \frac{\beta}{1-\beta}\,,
    \label{definition_omega_G}
\end{equation}
and
$\omega_{{\rm G},i} \coloneqq \omega_{\rm G} (u_i)$.
We refer to $\omega_{\rm G}$ as the source-driven frequency.
The last line of Eq.~\eqref{eq:spectrum_fandamental} represents the Fourier transform of a finite-duration source with a fixed complex frequency $\omega_{{\rm G},i}$, active over the interval $u_i \leq u_{\rm s} \leq u_{i+1}$.

Two properties of $\omega_{{\rm G}}$ will be important in the following analysis.
First, because $\beta<0$ during the plunge, $\text{Im} \ \omega_{{\rm G}}<0$, and hence $\omega_{{\rm G}}$ lies in the lower half of the complex-frequency plane.
Second, in the near-horizon limit,
\begin{equation}
    \omega_{{\rm G}} \to \omega_{\rm H}^{(1)} \ (= m\Omega_{\rm H}-i\kappa)\,,
\end{equation}
as $\beta \to -1$ and $\Omega \to \Omega_{\rm H}$, which coincides with the zero of the factor $(k_{\rm H} +i\kappa)$ in Eq.~\eqref{spectrum_final_form}.

To compare the two orders of integration, we write the segment spectrum as the product of a transfer function and a source factor.
The source factor contains the finite-duration window and the polynomial frequency dependence derived above.
This representation allows us either to evaluate the source-time integral first or to construct the time-domain response by convolution.

We define the transfer function as\footnote{The transfer function used by the pole-splitting method in Sec.~\ref{sec:Pole-Splitting-Method} includes an additional factor $B_{\ell m\omega}^{\rm out}$, compensated by its inverse in the effective source.
The ratio $B^{\rm out}_{\ell m\omega}/B^{\rm in}_{\ell m\omega}$ is the reflection amplitude of the curvature barrier, which becomes exponentially small once the barrier is transparent, so a weight carrying $1/B^{\rm out}_{\ell m\omega}$ grows at large real $\omega$.}
\begin{align}
\begin{split}
\label{eq:kernel}
  \mathcal{\widehat G}_{\ell m\omega}^\sigma \coloneqq\frac{\sigma(\omega)}{2i\omega B^{\rm in}_{\ell m\omega}}\,,
\end{split}
\end{align}
with $\sigma(\omega) \coloneqq S_{\ell m \omega}(\theta)S_{\ell m \omega}(\theta_0)$.\footnote{The factor $S_{\ell m\omega}(\theta_0)$ comes from the angular projection of the source, whereas $S_{\ell m\omega}(\theta)$ enters the reconstruction of $\Psi_4$ at the observation angle.
Their product is the angular factor in the separated Green's function~\cite{Su:2026gmp}.
}

We also define the following factors that constitute the integrand in Eq.~\eqref{eq:spectrum_fandamental}:
\begin{align}
\begin{split}
\label{eq:Wpoly}
\widetilde{\mathcal{W}}_i(\omega)&\coloneqq\mathcal{N}_iP(\omega)\,,\\
  P(\omega)&\coloneqq(\omega-\omega_{\rm H}^{(1)})(\omega-\omega_{\rm H}^{(2)})\,,\\
  \mathcal{N}_i&\coloneqq\mu\tilde{\mathcal{A}}
   \frac{4M\kappa e^{-2r_+\kappa}}{1-\beta_i}
   e^{-i\int^{u_i}\omega_{\rm G}du}\,.
\end{split}
\end{align}
Here $P$ is the screening factor in Eq.~\eqref{spectrum_final_form} represented by $\omega$, and $\mathcal{N}_i$ is a factor independent of $\omega$.
From Eqs.~\eqref{eq:kernel} and~\eqref{eq:Wpoly}, we have the segmented full spectrum as
\begin{align}
\begin{split}
\label{eq:zseg}
  z_{\ell m\omega[i]}S_{\ell m \omega}(\theta) &= e^{i\omega u_i}\mathcal{\widehat G}_{\ell m\omega}^{\sigma}\widetilde{\mathcal{W}}_i(\omega)\widetilde{\mathcal{I}}_i(\omega)\,,
\end{split}
\end{align}
where we defined
\begin{align}
\begin{split}
\label{eq:I_i_original}
    \widetilde{\mathcal{I}}_i(\omega) &\coloneqq \int_0^{\Delta u_i} d(\Delta u)e^{i(\omega-\omega_{{\rm G},i})\Delta u} \\
    &=\underbrace{\frac{e^{i(\omega-\omega_{{\rm G},i})\Delta u_i}}{i(\omega-\omega_{{\rm G},i})}}_{\eqqcolon \widetilde{\mathcal{I}}_i^{\rm A}(\omega)}\underbrace{-\frac{1}{i(\omega-\omega_{{\rm G},i})}}_{\eqqcolon \widetilde{\mathcal{I}}_i^{\rm B}(\omega)}\,,\\
\end{split}
\end{align}
which is the Fourier transform of the segment source windowed to $0\le\Delta u\le\Delta u_i$.
Then, the contribution of the $i$-th segment in $\Psi_{4,\ell m}$ is
\begin{align}
\begin{split}
\label{eq:Psiseg}
  \Psi_{4,\ell m[i]}(u)&=\frac{1}{r}\int\frac{d\omega}{2\pi}e^{-i\omega u'}\mathcal{\widehat G}_{\ell m\omega}^\sigma\widetilde{\mathcal{W}}_i(\omega)\widetilde{\mathcal{I}}_i(\omega)\\
  &=\frac{1}{r}\int\frac{d\omega}{2\pi}\int_0^{\Delta u_i}d(\Delta u)e^{-i\omega (u'-\Delta u)}\\
  &\qquad\times e^{-i\omega_{{\rm G},i}\Delta u} \mathcal{\widehat G}_{\ell m\omega}^\sigma\widetilde{\mathcal{W}}_i(\omega)\,.
\end{split}
\end{align}
We define $u'\coloneqq u-u_i$ as the observer's retarded-time offset from the beginning of the source segment.
Unlike the source integration variable $\Delta u$, $u'$ is fixed when evaluating the waveform at a given observation time.
The combination $u'-\Delta u=u-u_{\rm s}(t)$ in Eq.~\eqref{eq:Psiseg} is the time argument of the Green's function.
We define the reduced waveform as $\psi_i(u)\coloneqq r \Psi_{4,\ell m[i]}(u)$, and omit the subscripts $\ell m$ hereinafter.

The analytic structure of the transfer function determines the contributions obtained when the frequency contour is deformed.
The transfer function has three kinds of poles or singularities:
(i) The zeros of $B^{\rm in}_{\omega}$ are the QNM frequencies $\omega_n$
, at which $\mathcal{\widehat G}^{\sigma}_{\omega}$ has simple poles with residues defined as
\begin{equation}
\label{eq:Esigma}
  \mathsf{E}^{\sigma}_n\coloneqq \left.\frac{\sigma(\omega)}{2i\omega\,(dB^{\rm in}_{\omega}/d\omega)}\right|_{\omega=\omega_n}\,.
\end{equation}
(ii) A branch point sits at $\omega=0$, and the cut is conventionally introduced on the negative imaginary axis, which leads to the power-law tail at late times~\cite{Leaver:1986gd}.
(iii) Ref.~\cite{Su:2026gmp} showed that there are also square-root branch points starting from the exceptional points where two angular eigenvalues coalesce.
These points are distributed symmetrically about the real axis.
They have no counterpart in the Schwarzschild case.
In the analytic contour calculations below, we ignore the angular-cut jump integrals, following the prescription of Ref.~\cite{Su:2026gmp}.

The frequency integral in Eq.~\eqref{eq:Psiseg} has another pole\footnote{
Although the function $\widetilde{\mathcal{I}}_i (\omega)$ is regular at $\omega = \omega_{\rm G,i}$, the first and second terms in Eq.~\eqref{eq:I_i_original} have different contours in the $\omega$-integration in Eq.~\eqref{eq:Psiseg}, and the poles lead to a non-zero contribution only when $0 < u' < \Delta u_i$.} at $\omega=\omega_{{\rm G},i}$, which leads to the DW in the time-domain signal.

\begin{figure*}[t]
  \centering
  \includegraphics[width=\textwidth]{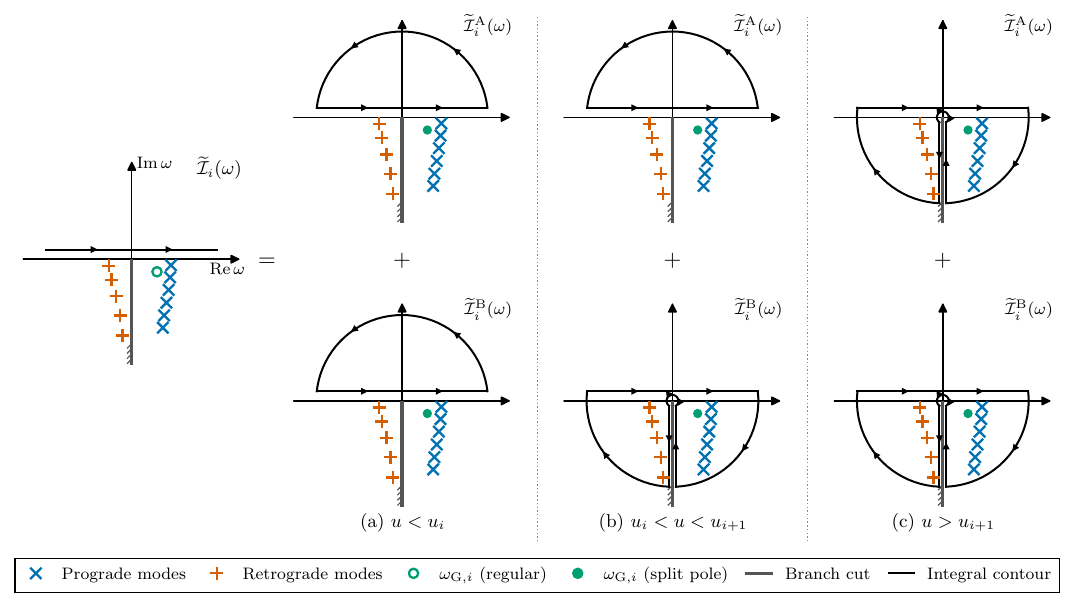}
  \caption{Contours used to evaluate the segment integral of
    Eq.~\eqref{eq:Psiseg}.
    The marked singularities are those of the transfer function $\mathcal{\widehat G}^{\sigma}_{\omega}$, and each panel is labelled by the term of Eq.~\eqref{eq:I_i_original}.
    The left panel is the undeformed contour, on which $\omega_{{\rm G},i}$ is a regular point (open circle).
    The three groups on the right are the same splitting in the three observation windows, with term A above and term B below, and each term acquires a simple pole at $\omega_{{\rm G},i}$ (filled circle).
    Term A carries $e^{-i\omega(u-u_{i+1})}$ and term B carries $e^{-i\omega(u-u_i)}$, so each closes in the half-plane fixed by the sign of its exponent, the upper closures vanishing under the prescription of Sec.~\ref{sec:source}.
    In (a), both close above and $\psi_i=0$.
    In (b), term B closes below; its residue at $\omega_{{\rm G},i}$ gives the DW of Eq.~\eqref{eq:DW}, and the singularities of $\mathcal{\widehat G}^{\sigma}_{\omega}$ give
    Eqs.~\eqref{eq:m2-modQNM}--\eqref{eq:m2-modprompt}.
    In (c), both contours are closed below and enclose $\omega_{{\rm G},i}$; the two source-driven frequency residues cancel, leaving contributions from the QNM poles, the branch cut, and the large-arc.
    Crosses are the prograde frequencies $\omega_{\ell m n}$ and plus signs the retrograde ones $-\bar\omega_{\ell -m n}$}
  \label{fig_contour_image}
\end{figure*}

%%%%%%%%%%%%%%%%%%%%%%%%%%%%
\subsection{DW from the frequency-domain Green's function method}
\label{subsec:dw_from_spectral}
%%%%%%%%%%%%%%%%%%%%%%%%%%%%
We use the frequency-domain method to isolate the term oscillating at the source-driven frequency and determine its amplitude.
Starting from Eq.~\eqref{eq:Psiseg}, we evaluate the source-time integral and deform the frequency contours for the two endpoint terms separately.
This procedure yields the DW contribution defined below.

The different exponents in Eq.~\eqref{eq:I_i_original} affect the convergence of the $\omega$ integral in Eq.~\eqref{eq:Psiseg} when deforming the frequency contour.
The split introduces an artificial simple pole at $\omega_{{\rm G},i}$ in each term individually in Eq.~\eqref{eq:I_i_original}.
In the window $u_i<u<u_{i+1}$ one has $u'-\Delta u_i<0$, so term A with $e^{-i\omega(u'-\Delta u_i)}$ is closed in the upper half-plane, where it vanishes.\footnote{This step is where the angular branch points of Sec.~\ref{sec:source} enter.
For a fixed $(\ell,m)$ mode, no horizontal contour at finite ${\rm Im}\,\omega$ lies above all of them, so the standard mode-by-mode argument for the vanishing of the response before the arrival of the signal does not apply~\cite{Su:2026gmp}.
The obstruction belongs to the separated representation, since summing over $\ell$ at fixed $m$ cancels the angular cuts.
We work mode by mode without using that cancellation.
The statements below hold to the accuracy to which these integrals may be neglected, and we do not estimate their size.
}
In the same time window, term B with $e^{-i\omega u'}$ is closed in the lower half-plane.
It collects the residue at the split pole $\omega_{{\rm G},i}$ of Eq.~\eqref{eq:I_i_original}, which gives
\begin{align}
\begin{split}
\label{eq:DW}
  \psi_{i}^{\rm DW}(u)=
  \mathcal{\widehat G}_{\omega_{{\rm G},i}}^\sigma\widetilde{\mathcal{W}}_i(\omega_{{\rm G},i})e^{-i\omega_{{\rm G},i}u'}\,.
\end{split}
\end{align}
We define the DW by Eq.~\eqref{eq:DW} within the segment window.

The remaining contributions come from the QNM poles, the radial branch cut, and the large-arc,
\begin{align}
\begin{split}
\label{eq:m2-modQNM}
  \psi^{\rm mod\text{-}QNM}_i=
  -\sum_n\frac{\mathsf{E}_n^\sigma\widetilde{\mathcal{W}}_i(\omega_n)}{\omega_{{\rm G},i}-\omega_n}e^{-i\omega_n u'}\,,
\end{split}
\end{align}
\begin{align}
\begin{split}
\label{eq:m2-modtail}
  \psi^{\rm mod\text{-}tail}_i=
  -\int_{\mathcal{C}_{\rm bc}}\frac{d\omega}{2\pi i}\frac{\mathcal{\widehat G}_{\omega}^\sigma\widetilde{\mathcal{W}}_i(\omega)}{\omega-\omega_{{\rm G},i}}e^{-i\omega u'}\,,
\end{split}
\end{align}
\begin{align}
\begin{split}
\label{eq:m2-modprompt}
  \psi^{\rm mod\text{-}prompt}_i=
  -\int_{\mathcal{C}_{\rm arc}}\frac{d\omega}{2\pi i}\frac{\mathcal{\widehat G}_{\omega}^\sigma\widetilde{\mathcal{W}}_i(\omega)}{\omega-\omega_{{\rm G},i}}e^{-i\omega u'}\,.
\end{split}
\end{align}
The contours $\mathcal{C}_{\rm bc}$ and $\mathcal{C}_{\rm arc}$ are the branch-cut and large-arc portions of the closed contour of FIG.~\ref{fig_contour_image}, both closed clockwise.

The DW amplitude is obtained by evaluating the transfer function and source weight at the segment frequency $\omega_{{\rm G},i}$.
Restoring the factor $1/r$ and the explicit source dependence gives
\begin{align}
\begin{split}
\label{eq:DW_extraction}
\Psi_{4 [i]}^{\rm DW} &=  \frac{1}{r}\left[ \frac{\mu \tilde{\cal A}S_{\omega} (\theta)S_{\omega} (\theta_0)} {i\omega B_{\omega}^{\rm in}}\left(k_{\rm H} + i \kappa \right) \left(k_{\rm H} +2 i \kappa \right) \right]_{\omega = \omega_{{\rm G},i}}\\
    &\times \frac{2M \kappa e^{-2 r_+ \kappa}}{1-\beta_i} \exp \left[-i \int^{u_i}  \omega_{\rm G} du \right]\\
    &\times \exp \left[ -i \omega_{{\rm G}, i} (u- u_i) \right]\,,
    \end{split}
    \end{align}
for $u_i < u < u_{i+1}$, and $\Psi_{4,[i]}^{\rm DW} = 0$ otherwise.

We now combine the local DW terms from successive source segments.
Taking $\Delta u_i$ short compared with the timescale on which $\omega_{\rm G}$ evolves gives $\exp[-i\int^{u_i}\omega_{\rm G}du]\,e^{-i\omega_{{\rm G},i}(u-u_i)}\to\exp[-i\int^{u}\omega_{\rm G}du]$, and hence
\begin{align}
\begin{split}
\Psi_{4}^{\rm DW} &=  \frac{1}{r}\left[ \frac{\mu \tilde{\cal A} S_{\omega} (\theta) S_{\omega} (\theta_0)} {i\omega B_{\omega}^{\rm in}}\left(k_{\rm H} + i \kappa \right) \left(k_{\rm H} +2 i \kappa \right) \right]_{\omega = \omega_{\rm G}}\\
    &\times \frac{2M \kappa e^{-2r_+ \kappa}}{1-\beta} \exp \left[-i \int^{u}  \omega_{\rm G} du \right]\,.
    \end{split}
    \label{psi_4_dw_approx}
\end{align}
This expression gives the DW contribution along the trajectory.
The exponential contains the frequency evolution associated with the particle motion, while the prefactor contains the frequency dependence of the source weight and the transfer function.

%%%%%%%%%%%%%%%%%%%%%%%%%%%%
\subsection{DW from the time-domain Green's function method}
\label{sec:DW-analytic}
%%%%%%%%%%%%%%%%%%%%%%%%%%%%
We now examine how an oscillation at the source-driven frequency arises from the QNM part of the Green's function.
We first transform the transfer function and source factor to the time domain and then evaluate their convolution.
This calculation identifies the term oscillating at the source-driven frequency within the QNM response without introducing a pole at $\omega_{{\rm G},i}$.

Let us first consider the inverse Fourier transform of the transfer function of Eq.~\eqref{eq:kernel},
\begin{align}
\begin{split}
\label{eq:Gdef}
  G(s)&=\int\frac{d\omega}{2\pi}e^{-i\omega s}\mathcal{\widehat G}_{\omega}^\sigma\,.
\end{split}
\end{align}
Closing the contour of Eq.~\eqref{eq:Gdef} in the lower-half of the complex-frequency plane for $s>0$ gives Leaver's decomposition~\cite{Leaver:1986gd},
\begin{align}
\begin{split}
\label{eq:G-decomp}
    G(s)&=G^{\rm QNM}(s) + G^{\rm cut}(s)+G^{\rm arc}(s)\,, \\
    G^{\rm QNM}(s)&=-i\sum_n\mathsf{E}_n^\sigma e^{-i\omega_n s}\Theta(s)\,.
\end{split}
\end{align}
Here $\Theta$ denotes the Heaviside step function.
Closing the contour in the upper-half-frequency plane for $s<0$ gives $G(s)=0$ under the prescription of Sec.~\ref{sec:source}.

We next perform the inverse transform of the source.
Equation~\eqref{eq:I_i_original} defines $\widetilde{\mathcal{I}}_i$ as the Fourier transform of the windowed function $I_i(\Delta u)$
\begin{align}
\begin{split}
\label{eq:window}
  I_i(\Delta u)\coloneqq e^{-i\omega_{{\rm G},i}\Delta u}\Theta(\Delta u)\Theta(\Delta u_i - \Delta u)\,,
\end{split}
\end{align}
in the convention $f(\Delta u)=\int\frac{d\omega}{2\pi}e^{-i\omega\Delta u}\tilde f(\omega)$ used throughout.
In the time domain, the function $\widetilde{\mathcal{W}}_i$ in Eq.~\eqref{eq:Wpoly} acts on $I_i$ as a differential operator.
From $\omega\,e^{-i\omega\Delta u}=i\partial_{\Delta u} e^{-i\omega\Delta u}$, multiplication by $\widetilde{\mathcal{W}}_i(\omega)$ becomes the operator $\widetilde{\mathcal{W}}_i(i\partial_{\Delta u})$.
The window in Eq.~\eqref{eq:window} is discontinuous at both ends, so the derivatives produce delta functions there.
The time-domain source is
\begin{align}
\begin{split}
\label{eq:Stime}
  S_i(\Delta u)&\coloneqq \int\frac{d\omega}{2\pi}e^{-i\omega\Delta u}\widetilde{\mathcal{W}}_i(\omega)\widetilde{\mathcal{I}}_i(\omega) \\
  &=\widetilde{\mathcal{W}}_i(\omega_{{\rm G},i})I_i(\Delta u)+\mathcal{N}_i\Bigl[i(\omega_{{\rm G},i}-\omega_{\rm H}^{(1)}-\omega_{\rm H}^{(2)})\\
  &\qquad\times\bigl(\delta(\Delta u)-e^{-i\omega_{{\rm G},i}\Delta u_i}\delta(\Delta u-\Delta u_i)\bigr)\\
  &\qquad-\bigl(\delta'(\Delta u)-e^{-i\omega_{{\rm G},i}\Delta u_i}\delta'(\Delta u-\Delta u_i)\bigr)\Bigr]\,.
\end{split}
\end{align}
The first term describes the source in the interior of the segment, with the constant amplitude $\widetilde{\mathcal{W}}_i(\omega_{{\rm G},i})$.
The remaining terms sit at the two boundaries $\Delta u=0$ and $\Delta u=\Delta u_i$ of the window.

With the time-domain Green's function and the source, we can calculate the full waveform.
The segment response is the retarded convolution of Eqs.~\eqref{eq:Gdef} and \eqref{eq:Stime},
\begin{align}
\begin{split}
\label{eq:convolution}
  \psi_i(u)&=\int_{-\infty}^{\infty}d(\Delta u) G(u'-\Delta u)S_i(\Delta u)\,.
\end{split}
\end{align}
In this convolution, $u'$ is the fixed observation-time offset and $\Delta u$ is the source-time integration variable.
Under the retarded prescription adopted above, the Green's function restricts the source contribution to $\Delta u\leq u'$.
For $0<u'<\Delta u_i$, the interior source integral therefore extends from $\Delta u=0$ to $\Delta u=u'$, while the boundary terms at $\Delta u=\Delta u_i$ do not contribute.
This is the time-domain counterpart of term A vanishing in Sec.~\ref{subsec:dw_from_spectral}.
With $\int d\Delta u\delta(\Delta u)G(u'-\Delta u)=G(u')$ and $\int d\Delta u\delta'(\Delta u)G(u'-\Delta u)=\dot G(u')$, we get
\begin{align}
\begin{split}
\label{eq:psi-split}
  \psi_i(u)&=\widetilde{\mathcal{W}}_i(\omega_{{\rm G},i})
    \int_0^{u'}d(\Delta u) e^{-i\omega_{{\rm G},i}\Delta u}G(u'-\Delta u)\\
  &+\mathcal{N}_i\Bigl[i(\omega_{{\rm G},i}-\omega_{\rm H}^{(1)}-\omega_{\rm H}^{(2)})G(u')-\dot G(u')\Bigr]\,,
\end{split}
\end{align}
where the dot denotes the derivative with respect to the argument of $G$.

The convolution with the QNM Green's function produces terms at both the source-driven frequency and the QNM frequencies.
To see the contribution of QNM poles in $\psi_i$, let us substitute $G^{\rm QNM}$ in the first line of Eq.~\eqref{eq:psi-split}.
The integral is then elementary and gives
\begin{align}
\begin{split}
\label{eq:drive-integral}
  \widetilde{\mathcal{W}}_i(\omega_{{\rm G},i})\sum_n\frac{\mathsf{E}_n^\sigma}{\omega_{{\rm G},i}-\omega_n}
  \Bigl(e^{-i\omega_{{\rm G},i}u'}-e^{-i\omega_n u'}\Bigr)\,.
\end{split}
\end{align}
The upper integration limit produces the exponential at $\omega_{{\rm G},i}$, whereas the lower limit produces the exponentials at $\omega_n$.
The term oscillating at the source-driven frequency therefore arises directly from the convolution with the QNM Green's function.

We evaluate the second line of Eq.~\eqref{eq:psi-split} next.
Equation~\eqref{eq:G-decomp} gives $\dot G^{\rm QNM}(u')=-\sum_n\mathsf{E}_n^\sigma\omega_n e^{-i\omega_n u'}$ at $u'>0$, so the second line becomes
\begin{align}
\begin{split}
\label{eq:contact-raw}
  &\mathcal{N}_i\sum_n\mathsf{E}_n^\sigma
  \bigl(\omega_{{\rm G},i}+\omega_n-\omega_{\rm H}^{(1)}-\omega_{\rm H}^{(2)}\bigr)e^{-i\omega_n u'}\\
  &=\sum_n\mathsf{E}_n^\sigma\widetilde{\mathcal{W}}_i[\omega_n,\omega_{{\rm G},i}]e^{-i\omega_n u'}\,,
\end{split}
\end{align}
where we defined
\begin{equation}
\label{eq:Wdd}
    \widetilde{\mathcal{W}}_i[\alpha,\beta]\coloneqq\frac{\widetilde{\mathcal{W}}_i(\beta)-\widetilde{\mathcal{W}}_i(\alpha)}{\beta-\alpha}\,.
\end{equation}
Adding Eq.~\eqref{eq:contact-raw} to Eq.~\eqref{eq:drive-integral} replaces $\widetilde{\mathcal{W}}_i(\omega_{{\rm G},i})$ by $\widetilde{\mathcal{W}}_i(\omega_n)$ in the coefficients of $e^{-i\omega_n u'}$,
\begin{align}
\begin{split}
\label{eq:psi-pole}
  \psi_i^{\rm QNM}(u)&=\widetilde{\mathcal{W}}_i(\omega_{{\rm G},i})\sum_n \frac{\mathsf{E}_n^\sigma}{\omega_{{\rm G},i}-\omega_n}e^{-i\omega_{{\rm G},i}u'}\\
     &-\sum_n\frac{\mathsf{E}_n^\sigma\widetilde{\mathcal{W}}_i(\omega_n)}{\omega_{{\rm G},i}-\omega_n}e^{-i\omega_n u'}\,,
\end{split}
\end{align}
The QNM-frequency terms in the second line agree with Eq.~\eqref{eq:m2-modQNM}.
We rewrite Eq.~\eqref{eq:psi-pole} in terms of QNM exponentials with time-dependent amplitudes:
\begin{align}
\label{eq:psi-dynamical}
  \psi_i^{\rm QNM}(u)=\sum_n \mathcal{C}_{n,i}(u')\,e^{-i\omega_n u'}\,,
\end{align}
where
\begin{align}
\label{eq:Ccoef}
  \mathcal{C}_{n,i}(u')\coloneqq\frac{\mathsf{E}_n^\sigma}{\omega_{{\rm G},i}-\omega_n}
  \Bigl[\widetilde{\mathcal{W}}_i(\omega_{{\rm G},i})e^{-i(\omega_{{\rm G},i}-\omega_n)u'}-\widetilde{\mathcal{W}}_i(\omega_n)\Bigr]\,.
\end{align}
Equation~\eqref{eq:psi-dynamical} expresses the QNM response in terms of time-dependent amplitudes $\mathcal{C}_{n,i}$.
These amplitudes encode the source-driven frequency dependence, giving the form of dynamical QNM excitation~\cite{DeAmicis:2025xuh}.
%

%%%%%%%%%%%%%%%%%%%%%%%%%%%%
\subsection{Decomposition of the direct-wave amplitude}
\label{subsec:DW-composition}
%%%%%%%%%%%%%%%%%%%%%%%%%%%%

Here, we will relate the calculations of the source-driven terms obtained in Sec.~\ref{subsec:dw_from_spectral} and Sec.~\ref{sec:DW-analytic}, and discuss how the DW is understood in the response.
As we did in Sec.~\ref{sec:DW-analytic}, dividing the transfer function into the QNM sector and the rest can facilitate understanding of the waveform components.
While keeping the integration technique to obtain the waveform the same as that of Sec.~\ref{subsec:dw_from_spectral}, we separate the transfer function of Eq.~\eqref{eq:kernel} into its QNM sector and the remainder,
\begin{align}
\begin{split}
\label{eq:Dsplit-kernel}
  \mathcal{\widehat G}_{\omega}^\sigma&=\underbrace{\sum_{n}\frac{\mathsf{E}_n^\sigma}{\omega-\omega_n}}_{\eqqcolon \mathcal{\widehat G}_{\rm pole}^\sigma(\omega)}+ \underbrace{\left\lbrack \mathcal{\widehat G}_{\omega}^\sigma - \sum_{n}\frac{\mathsf{E}_n^\sigma}{\omega-\omega_n}\right\rbrack}_{\eqqcolon \mathcal{\widehat G}_{\rm non}^\sigma(\omega)}\,,\\
\end{split}
\end{align}
with the sum running over all the poles.\footnote{We define $\mathcal{\widehat G}^{\sigma}_{\rm non}$ by the subtraction in Eq.~\eqref{eq:Dsplit-kernel}, and we do not address the convergence of the pole sum here.
The numerical implementation truncates it at a finite overtone number, as described in Appendix~\ref{app_detail_PSmethod}.
This is the splitting that Sec.~\ref{sec:Pole-Splitting-Method} applies to the waveform.}

The two terms of Eq.~\eqref{eq:Dsplit-kernel} carry different singularities by construction.
The singularities of $\mathcal{\widehat G}^{\sigma}_{\rm pole}$ are the simple poles at $\omega_n$, and $\mathcal{\widehat G}^{\sigma}_{\rm non}$ carries the branch cut running from the origin and the angular cuts listed in Sec.~\ref{sec:source}.
The pole sector is regular in the upper half-plane, and the non-pole sector is treated under the prescription of Sec.~\ref{sec:source}.

We organize the components of the waveform when the decomposition in Eq.~\eqref{eq:Dsplit-kernel} is applied.
Applying this decomposition to the segment spectrum gives $\psi_i=\psi_i^{\rm pole}+\psi_i^{\rm non}$, where
\begin{align}
\begin{split}
\label{eq:sector-split}
  \psi_i^{\rm pole/non}(u)=\int\frac{d\omega}{2\pi}e^{-i\omega u'}
    \mathcal{\widehat G}_{\rm pole/non}^\sigma(\omega)\widetilde{\mathcal{W}}_i(\omega)\widetilde{\mathcal{I}}_i(\omega)\,.
\end{split}
\end{align}
We follow the same integration procedure as in Sec.~\ref{subsec:dw_from_spectral} to obtain the waveform, so that the term B in Eq.~\eqref{eq:I_i_original} gives all components in each sector as
\begin{align}
\begin{split}
\label{eq:psi-pole-sec}
    \psi_i^{\rm pole}(u)&=\mathcal{\widehat G}_{\rm pole}^\sigma(\omega_{{\rm G},i})\widetilde{\mathcal{W}}_i(\omega_{{\rm G},i})e^{-i\omega_{{\rm G},i}u'}\\
    &-\sum_n\frac{\mathsf{E}_n^\sigma\widetilde{\mathcal{W}}_i(\omega_n)}{\omega_{{\rm G},i}-\omega_n}e^{-i\omega_n u'}\\
    &-\int_{\mathcal{C}_{\rm arc}}\frac{d\omega}{2\pi i}\frac{\mathcal{\widehat G}^{\sigma}_{\rm pole}(\omega)\widetilde{\mathcal{W}}_i(\omega)}{\omega-\omega_{{\rm G},i}}e^{-i\omega u'}\,,
\end{split}
\end{align}
and
\begin{align}
\begin{split}
\label{eq:psi-non-sec}
    \psi_i^{\rm non}(u)&=\mathcal{\widehat G}_{\rm non}^\sigma(\omega_{{\rm G},i})\widetilde{\mathcal{W}}_i(\omega_{{\rm G},i})e^{-i\omega_{{\rm G},i}u'}\\
    &-\Big[\int_{\mathcal{C}_{\rm bc}}+\int_{\mathcal{C}_{\rm arc}}\Big]\frac{d\omega}{2\pi i}\frac{\mathcal{\widehat G}_{\rm non}^\sigma(\omega)\widetilde{\mathcal{W}}_i(\omega)}{\omega-\omega_{{\rm G},i}}e^{-i\omega u'}\,.
\end{split}
\end{align}
The first lines of Eqs.~\eqref{eq:psi-pole-sec} and ~\eqref{eq:psi-non-sec} sum to Eq.~\eqref{eq:DW}, because $\mathcal{\widehat G}_{\rm pole}^\sigma+\mathcal{\widehat G}_{\rm non}^\sigma=\mathcal{\widehat G}_{\omega}^\sigma$ by construction, which reads
\begin{align}
\begin{split}
\label{eq:DW-shares}
  \psi_i^{\rm DW}=\Big[\mathcal{\widehat G}_{\rm pole}^\sigma(\omega_{{\rm G},i})+\mathcal{\widehat G}^\sigma_{\rm non}(\omega_{{\rm G},i})\Big]
  \widetilde{\mathcal{W}}_i(\omega_{{\rm G},i})e^{-i\omega_{{\rm G},i}u'}\,.
\end{split}
\end{align}

The pole-sector contribution to the DW is contained in the dynamically excited QNM response.
The first term of Eq.~\eqref{eq:DW-shares} is exactly the first line of Eq.~\eqref{eq:psi-pole} in Sec.~\ref{sec:DW-analytic} because of the following identity,
\begin{align}
\begin{split}
\label{eq:pole-share-m1}
  \mathcal{\widehat G}_{\rm pole}^\sigma(\omega_{{\rm G},i})=\sum_n\frac{\mathsf{E}_n^\sigma}{\omega_{{\rm G},i}-\omega_n}\,.
\end{split}
\end{align}
The first term of Eq.~\eqref{eq:DW-shares} turns out to be a part of the dynamical QNM excitation in Ref.~\cite{DeAmicis:2025xuh}.
Eq.~\eqref{eq:Ccoef} incorporates this same source-driven frequency term into the time-dependent QNM amplitudes.

We now examine which sector can be used to trace the source-driven frequency $\omega_{\rm G}$.
The pole sector in Eq.~\eqref{eq:psi-pole-sec} contains both the exponential at $\omega_{{\rm G},i}$ and terms at the QNM frequencies $\omega_n$, whose interference generally causes its instantaneous frequency to differ from $\omega_{{\rm G},i}$.
In contrast, the non-pole sector in Eq.~\eqref{eq:psi-non-sec} contains the exponential at $\omega_{{\rm G},i}$ and the branch-cut and large-arc integrals, but no QNM terms.
We leave these integrals unevaluated and assume that their contributions to the waveform and its time derivative are negligible over the interval of interest.
Under this approximation, the non-pole response within each source segment is dominated by the exponential at $\omega_{{\rm G},i}$.
The non-pole sector is therefore expected to reveal the frequency evolution associated with the source without interference from the explicit QNM terms.
%

%%%%%%%%%%%%%%%%%%%%%%%%%%%%
\subsection{Near-horizon cancellation and its relation to the DW}
\label{sec:NH-cancellation}
%%%%%%%%%%%%%%%%%%%%%%%%%%%%

The preceding subsections relate the DW to dynamical QNM excitation and identify the non-pole sector as a candidate for tracing the source-driven frequency $\omega_{\rm G}$.
Since $\omega_{\rm G}$ approaches the fundamental horizon-mode frequency as the particle approaches the horizon, let us clarify how this DW description is related to the horizon-mode cancellation ~\cite{Zimmerman:2011dx,Kubota:2026qjh}.
We show that the cancellation of source-induced horizon-mode poles at $\omega=\omega_{\rm H}^{(n)}=m\Omega_{\rm H}-in\kappa$, with $n=1,2,\ldots$, persists when the point-particle trajectory, the ingoing radial solution $R^{\rm in}$, and the source coefficients are consistently expanded to all orders in $\epsilon=r-r_+$, which are partially neglected in Ref.~\cite{Kubota:2026qjh}.
This implies that if DWs emitted by a point particle are observable, they must originate from a region where the near-horizon expansion of the geodesic motion is not applicable. 
We will examine this possibility in Sec.~\ref{sec:Numerical-direct-wave}.

To complete our analysis, one has to examine the contribution of $R^{\rm in}$ as its $\epsilon$-expansion has poles at the horizon-mode frequencies~\cite{Motohashi:2026mbn}.
Also, we should carefully treat the source coefficients $A_{ab}$ up to higher orders as they may play an important role in the cancellation of the horizon modes (for the leading-order cancellation, see Ref.~\cite{Zimmerman:2011dx}).
Our calculation shows that the zeros from the source term and $1/B^{\rm in}$ continue to cancel the horizon-mode poles when we include these higher-order radial and source contributions.

We perform this analysis in the extreme-mass-ratio limit for a geodesic point particle, neglecting radiation reaction and other finite-mass-ratio effects.
Whether the cancellation persists beyond these approximations remains to be investigated.

Let us expand the spectral amplitude $Z_{\ell m \omega}$ \eqref{spectrum_original_for_far} in $\epsilon=r-r_+$:
\begin{align}
Z_{\ell m \omega} = \frac{1}{2i \omega B_{\ell m \omega}^{\rm in}} \int_{r_+}^{\infty} dr R_{\ell m \omega}^{\rm in} \Delta^{-2} T_{\ell m \omega}\,,
\end{align}
and investigate whether it has horizon-mode poles at the $n$-th-order expansion coefficient.
The analysis for the leading-order term was already performed by Zimmerman and Chen \cite{Zimmerman:2011dx}.

For fixed $(\ell,m)$, let us introduce $q=i(\omega-m\Omega_{\rm H})/\kappa$, so that $\omega(q)=m\Omega_{\rm H}-i\kappa q$ and $q=j$ ($j = 1, 2, 3, ...$) corresponds to $\omega_{\rm H}^{(j)}$.
In the following, we omit the angular indices, $\ell$ and $m$, and write $R^{\rm in}(\epsilon,q)=R^{\rm in}_{\ell m\omega(q)}(r_++\epsilon)$ and $B^{\rm in}(q)=B^{\rm in}_{\ell m\omega(q)}$, with the normalization of Eq.~\eqref{eq:Rin_asympt}.
We expand the ingoing solution, $R^{\rm in}$, around the horizon as
\begin{align}
    R^{\rm in}(\epsilon,q)=\epsilon^{2-q/2}\sum_{n=0}^{\infty}c_n(q)\epsilon^n,
\label{eq:NH-Frobenius}
\end{align}
where $c_0$ is an analytic and non-zero coefficient.
The coefficients $c_n$ ($n \geq 1$) have at most simple poles at $q=3,\ldots,n+2$~\cite{Motohashi:2026mbn}.\footnote{We consider $\kappa>0$ and a horizon-crossing timelike geodesic with $E-\Omega_{\rm H}L_z>0$.
The angular eigenvalue and harmonic are assumed analytic near the frequency under consideration.
Frequency continuation is the same as in the retarded Green's function.}
For the connection coefficient $B^{\rm in}$ with the spin weight of $s=-2$, we assume that it is analytic and nonzero at $q=1$ and $2$, and has a simple pole with nonzero residue at each $q=j\ge3$ \cite{Keshet:2007be,Casals:2019vdb}.

Integrating the radial delta distributions and their derivatives in the point-particle source [see e.g., a review paper \cite{Sasaki:2003xr}] gives a source contraction with the radial solution.
Retaining all six source coefficients, we define this contraction for a radial function $R$ by
\begin{align}
\begin{split}
    T[R]=&(A_{nn0}+A_{\bar mn0}+A_{\bar m\bar m0})R\\
    &-(A_{\bar mn1}+A_{\bar m\bar m1})R'+A_{\bar m\bar m2}R''\,,
\end{split}
\label{eq:NH-full-source}
\end{align}
where the coefficients are those of Ref.~\cite{Sasaki:2003xr}, with the particle mass factored out, and primes denoting radial derivatives.
With the contraction in Eq.~\eqref{eq:NH-full-source} evaluated along the trajectory, the complete-source version of Eq.~\eqref{spectra_mm_form} is
\begin{align}
\begin{split}
    Z_{\ell m\omega(q)}=&\frac{\mu}{2i\omega(q)B^{\rm in}(q)}\int dt e^{i\omega(q)t-im\phi(t)} T[R^{\rm in}](t,q)\,.
\end{split}
\label{eq:NH-full-spectrum}
\end{align}

To examine how the poles of the radial coefficients $c_n(q)$ enter the complete source, we use the full point-particle source coefficients~\cite{Sasaki:2003xr} to rewrite the contraction in Eq.~\eqref{eq:NH-full-source} in terms of the differential operator
\begin{equation}
    D_q=\partial_r-ip\,,
\label{eq:NH-Dq-definition}
\end{equation}
with $p=\frac{(r^2+a^2)\omega(q)-am}{\Delta}$.

We write $S=S_{\ell m\omega(q)}(\theta)$, $\varrho=(r-ia\cos\theta)^{-1}$, and $\bar\varrho=(r+ia\cos\theta)^{-1}$.
The angular differential operator is
\begin{equation}
    L_j^\dagger=\partial_\theta-\frac{m}{\sin\theta}+a\omega(q)\sin\theta+j\cot\theta\,,
\label{eq:NH-angular-operator}
\end{equation}
where $j=1,2$ labels the two operators.
The three factors $C_{nn}$, $C_{\bar mn}$, and $C_{\bar m\bar m}$ appearing in the source coefficients are obtained from the tetrad projections of the point-particle stress-energy tensor as
\begin{subequations}
\begin{align}
\label{eq:NH-tetrad-coefficients}
    C_{nn}&=\frac{1}{4\Sigma^3 u^t}\left[E(r^2+a^2)-aL_z+\Sigma\frac{dr}{d\tau}\right]^2\,,\\
    C_{\bar mn}&=-\frac{\varrho}{2\sqrt{2}\Sigma^2 u^t}\left[E(r^2+a^2)-aL_z+\Sigma\frac{dr}{d\tau}\right]\notag\\
    &\times\left[i\sin\theta\left(aE-\frac{L_z}{\sin^2\theta}\right)+\Sigma\frac{d\theta}{d\tau}\right]\,,\\
    C_{\bar m\bar m}&=\frac{\varrho^2}{2\Sigma u^t}\left[i\sin\theta\left(aE-\frac{L_z}{\sin^2\theta}\right)+\Sigma\frac{d\theta}{d\tau}\right]^2\,.
\end{align}
\end{subequations}
Here $u^t=dt/d\tau$, $\tau$ is the particle's proper time, and $\Sigma$ is defined in Eq.~\eqref{eq:kerr-functions}.
With these definitions, the six source coefficients are
\begin{subequations}
\label{eq:NH-six-source-coefficients}
\begin{align}
    A_{nn0}&=-\frac{2C_{nn}}{\sqrt{2\pi}\Delta^2}\varrho^{-2}\bar\varrho^{-1}L_1^\dagger\!\left[\varrho^{-4}L_2^\dagger(\varrho^3S)\right]\,,\\
    A_{\bar mn0}&=\frac{2C_{\bar mn}\varrho^{-3}}{\sqrt{\pi}\Delta}\Bigl[(L_2^\dagger S)(ip+\varrho+\bar\varrho)\notag\\
    &-a\sin\theta\,S\,p(\bar\varrho-\varrho)\Bigr]\,,\\
    A_{\bar m\bar m0}&=-\frac{C_{\bar m\bar m}\varrho^{-3}\bar\varrho S}{\sqrt{2\pi}}\left[-ip'-p^2+2i\varrho p\right]\,,\\
    A_{\bar mn1}&=\frac{2C_{\bar mn}\varrho^{-3}}{\sqrt{\pi}\Delta}\Bigl[L_2^\dagger S+ia\sin\theta(\bar\varrho-\varrho)S\Bigr]\,,\\
    A_{\bar m\bar m1}&=-\frac{2C_{\bar m\bar m}\varrho^{-3}\bar\varrho S}{\sqrt{2\pi}}(ip+\varrho)\,,\\
    A_{\bar m\bar m2}&=-\frac{C_{\bar m\bar m}\varrho^{-3}\bar\varrho S}{\sqrt{2\pi}}\,.
\end{align}
\end{subequations}
Here $p'=\partial_r p$ at fixed $q$.

Substituting these coefficients into Eq.~\eqref{eq:NH-full-source} and collecting the radial derivatives into $D_q$ and $D_q^2$, we obtain
\begin{equation}
    T[R]=-F D_q^2R+2G D_qR+HR,
\label{eq:NH-source-operator}
\end{equation}
where $F=-A_{\bar m\bar m2}$, $G=\varrho F-A_{\bar mn1}/2$, and $H=A_{nn0}+A_{\bar mn0}-ipA_{\bar mn1}$.

The near-horizon behavior of the source term in Eq.~\eqref{eq:NH-source-operator} determines how the radial coefficients $c_n$ in Eq.~\eqref{eq:NH-Frobenius} enter $T[R^{\rm in}]$.
Along the geodesics of the particle, $F$, $G$, and $H$ can be expanded in $\epsilon$, so we write $F=f_1\epsilon+f_2\epsilon^2+\cdots$, $G=g_1\epsilon+\cdots$, and $H=h_1\epsilon+\cdots$.\footnote{The tetrad coefficients $C_{nn}$, $C_{\bar mn}$, and $C_{\bar m\bar m}$ are respectively $\Delta^3$, $\Delta^2$, and $\Delta$ times analytic functions along the geodesic.
In $H$, the terms proportional to $p$ cancel between $A_{\bar mn0}$ and $ipA_{\bar mn1}$.
Substitution of the source coefficients then gives the stated factorization without truncating the trajectory.}
Expanding $p=-iq/(2\epsilon)+p_{\rm reg}(\epsilon,q)$ with respect to $\epsilon$, we have
\begin{align}
    D_q=\partial_\epsilon-\frac{q}{2\epsilon}-ip_{\rm reg}(\epsilon,q)\,,
\label{eq:NH-derivative-operator}
\end{align}
where $p_{\rm reg}=p+iq/(2\epsilon)$ is analytic.
Applying the operator $D_q$  in Eq.~\eqref{eq:NH-derivative-operator} to the $n$-th order term of $R^{\rm in}$ in Eq.~\eqref{eq:NH-Frobenius} gives
\begin{align}
    D_q\epsilon^{2-q/2+n}=(n+2-q)\epsilon^{1-q/2+n}-ip_{\rm reg}\epsilon^{2-q/2+n}\,.
\label{eq:NH-first-derivative}
\end{align}
Applying $D_q$ once more yields
\begin{align}
\begin{split}
    D_q^2\epsilon^{2-q/2+n}=&(n+2-q)(n+1-q)\epsilon^{n-q/2}\\
    &-2ip_0(n+2-q)\epsilon^{1+n-q/2}+O(\epsilon^{2+n-q/2})\,,
\end{split}
\label{eq:NH-second-derivative}
\end{align}
where $p_0=p_{\rm reg}(0,q)$.
Substituting these into Eq.~\eqref{eq:NH-source-operator}, we find the $N$-th order expansion coefficient of the source term $T[R^{\rm in}]$, $C_N$, is
\begin{align}
    T[R^{\rm in}]=\epsilon^{1-q/2}\sum_{N=0}^{\infty}C_N(q)\epsilon^N\,,
\label{eq:NH-source-series}
\end{align}
with
\begin{align}
\begin{split}
    C_N=&-f_1(N+1-q)(N+2-q)c_N\\
    &+(N+1-q)\bigl[2if_1p_0+2g_1-f_2(N-q)\bigr]c_{N-1}\\
    &+\sum_{n=0}^{N-2}d_{Nn}(q)c_n\,.
\end{split}
\label{eq:NH-coefficient-structure}
\end{align}
At order $\epsilon^{N+1-q/2}$ in $T[R^{\rm in}]$, the term proportional to $c_N$ comes only from $-f_1\epsilon D_q^2R^{\rm in}$.
The factors $2if_1p_0-f_2(N-q)$ and $2g_1$ multiplying $(N+1-q)c_{N-1}$ come from $-FD_q^2R^{\rm in}$ and $2GD_qR^{\rm in}$, respectively.
The remaining contributions involve $c_n$ with $n\le N-2$ and are collected in $d_{Nn}(q)$, which are analytic in $q$.
The term $HR^{\rm in}$ contributes only to this sum, since $Hc_n\epsilon^{n+2-q/2}=O(\epsilon^{n+3-q/2})$.
For $N=0$, the term proportional to $c_{N-1}$ is omitted, and the sum is empty for $N=0,1$.

From Eq.~\eqref{eq:NH-coefficient-structure}, we find that $C_N$ has no poles at $q=N+1$ and $q=N+2$ due to the factors $(N+1-q) (N+2-q)$ in the first term and $(N+1-q)$ in the second term in Eq.~\eqref{eq:NH-coefficient-structure}.
Thus $C_N$ can have simple poles only at $q=3,4,\ldots,N-1,N$.
At the first two frequencies, Eq.~\eqref{eq:NH-coefficient-structure} also gives 
\begin{equation}
    C_0(q=1)=C_0(q=2)=C_1(q=2)=0\,.
    \label{CCC}
\end{equation}

To obtain the spectrum in Eq.~\eqref{eq:NH-full-spectrum}, multiply the source term in Eq.~\eqref{eq:NH-source-series} by the phase factor $e^{i \omega(q) t-im\phi(t)}$ and integrate over the source time.
Along the near-horizon plunge, $\epsilon=r(t)-r_+$ decreases monotonically so that we can use it as the integration variable.
The contribution from $0<\epsilon<\epsilon_0$ is therefore
\begin{align}
\begin{split}
    Z_{\rm NH}(q)=&\frac{\mu}{2i\omega(q)B^{\rm in}(q)}\int_0^{\epsilon_0}d\epsilon\left[-\frac{dt}{d\epsilon}\right]\\
    &\quad\times e^{i\omega(q)t(\epsilon)-im\phi(\epsilon)}T[R^{\rm in}](\epsilon,q)\,.
\end{split}
\label{eq:NH-source-integral}
\end{align}
Here $t(\epsilon)$ is the time along the orbit, $\phi(\epsilon)=\phi(t(\epsilon))$, and the remaining trajectory dependence is included in $T[R^{\rm in}]$.
We choose $\epsilon_0$ within the convergence region of the horizon expansion.

Near the horizon, the phase factor is $e^{i \omega(q) t-im\phi(t)} = \epsilon^{-q/2} \times \sum_{n=0}^{\infty} k_n \epsilon^n$ and $dt/d\epsilon = \epsilon^{-1} \times \sum_{n=0}^{\infty} l_n \epsilon^n$, where $k_n$ and $l_n$ are expansion coefficients.
We then expand the factors in the integrand as
\begin{align}
    \left[-\frac{dt}{d\epsilon}\right]e^{i\omega(q)t(\epsilon)-im\phi(\epsilon)}=\epsilon^{-1-q/2}\sum_{k=0}^{\infty}v_k(q)\epsilon^k\,,
\label{eq:NH-trajectory-series}
\end{align}
where $v_k(q)$ are expansion coefficients.
Multiplying the series in Eqs.~\eqref{eq:NH-source-series} and \eqref{eq:NH-trajectory-series} gives the integrand of Eq.~\eqref{eq:NH-source-integral} as $\epsilon^{-q}\sum_{N\ge0}A_N(q)\epsilon^N$, where the coefficients of the full integrand are
\begin{align}
    A_N(q)=\sum_{k=0}^{N}v_k(q)C_{N-k}(q)\,.
\label{eq:NH-integrand-coefficients}
\end{align}
Since the $v_k$ are analytic, $A_N$ has at most simple poles at $q=3,\ldots,N$ and satisfies $A_0(1)=A_0(2)=A_1(2)=0$.

Integrating the term proportional to $\epsilon^{N-q}$ with respect to $\epsilon$ leads to the factor $1/(N+1-q)$, which produces a candidate pole at $q=N+1$, or equivalently $\omega = \omega_{\rm H}^{(N+1)}$.
Term-by-term integration of Eq.~\eqref{eq:NH-source-integral} then gives
\begin{align}
    Z_{\rm NH}(q)=\frac{\mu}{2i\omega(q)}\sum_{N=0}^{\infty}\frac{\widehat A_N(q)\epsilon_0^{N+1-q}}{N+1-q}\,,
\label{eq:NH-amplitude-series}
\end{align}
where $\widehat A_N \coloneqq A_N/B^{\rm in}$.
At $q=N+1$, $\widehat A_N(q)$ vanishes.
For $N\ge2$, the zero originates from $1/B^{\rm in}$, and for $N=0,1$, the corresponding zeros originate from $A_0(1)=A_1(2)=0$ [see Eq.~\eqref{CCC}].
Consequently, for every $N\ge0$,
\begin{align}
    \widehat A_N(q= N+1)=0\,,
\label{eq:NH-coefficient-zero}
\end{align}
which cancels the poles at $q=N+1$ in Eq.~\eqref{eq:NH-amplitude-series}.
The possible poles of $A_N$ at $q=j$, with $3\le j\le N$, are canceled by the zeros of $1/B^{\rm in}$.

Under the stated analyticity and convergence conditions, these cancellations also ensure that the sum in Eq.~\eqref{eq:NH-amplitude-series} is analytic at the horizon-mode frequencies.\footnote{For a positive integer $j$, we choose $\epsilon_0$ within a common convergence domain of the normalized radial series and the analytic source and trajectory factors for $q$ near $j$.
Subtracting the terms with $N<j$ from $\sum_N\widehat A_N(q)\epsilon^N$ leaves a remainder uniformly $O(\epsilon^j)$.
Its contribution to the integral is analytic near $q=j$, since $\epsilon^{j-\operatorname{Re}q}$ is integrable for $\operatorname{Re}q<j+1$.
In the explicitly integrated finite sum, only the term with $N=j-1$ has a denominator $N+1-q$ that vanishes at $q=j$.
Its possible pole is canceled by the zero in Eq.~\eqref{eq:NH-coefficient-zero}.}
The cancellation of source-induced horizon-mode poles in $Z_{\rm NH}$ therefore persists to all orders in the near-horizon expansions of the radial solution, source, and trajectory.
The horizon-mode cancellation discussed by Kubota {\it et al.}~\cite{Kubota:2026qjh} therefore cannot be avoided by taking into account the higher orders of $\epsilon$ in the source coefficients, $A_{ab}$, as well as the function $R^{\rm in}$ under the idealized situations, e.g., extreme-mass-ratio approximations, ignoring the finite-size effect of a plunging object or deviations from the geodesic motion.

The DW calculation in the preceding subsections treats the orbital motion differently.
We use the leading near-horizon expressions for the radial solution and source coefficients, while Taylor-expanding $\phi(t)$ and $x(t)$ to first order about successive times $t_i$ over short intervals.
This local time expansion uses the angular and radial velocities at each point along the plunge and does not assume that the trajectory admits an all-order expansion solely in powers of $\epsilon=r-r_+ \propto e^{-2 \kappa t}$, as finite-mass-ratio or finite-size effects may introduce non-trivial corrections in the expansion.
Although we also assume the extreme-mass-ratio limit in this paper, our computation of $\omega_{\rm G}$ would be applicable to broader situations.
Also, an observable direct wave, whose frequency evolves as $\omega_{\rm G}$, may persist at earlier times, since the near-horizon expansion of the particle trajectory might not be valid near the ergosurface.
We have not quantitatively determined the domain of validity of the combined approximations or its relation to the convergence domain of the horizon expansion.
To numerically study the emission of observable direct waves, while avoiding the group delay caused by QNM filtering, we introduce the pole-splitting method in Sec.~\ref{sec:Pole-Splitting-Method} to analyze simulated waveforms.
We then assess the frequency model in Sec.~\ref{sec:Numerical-direct-wave} by comparing predictions based on $\omega_{\rm G}(u)$ with the instantaneous frequency of the numerical remainder.

%%%%%%%%%%%%%%%%%%%%%%%%%%%%
\section{The Pole-Splitting Method}
\label{sec:Pole-Splitting-Method}
%%%%%%%%%%%%%%%%%%%%%%%%%%%%

The analysis in Sec.~\ref{subsec:DW-composition} has suggested splitting the transfer function itself.
Unlike the rational or full QNM filtering used so far, which causes a non-trivial time shift in the resulting waveforms, the splitting can separate the time-domain waveform into the sum of the QNMs and the remainder, with the time axis untouched.
We define it here and work out its consequences for the DW numerically in Sec.~\ref{sec:Numerical-direct-wave}.

%%%%%%%%%%%%%%%%%%%%%%%%%%%%
\subsection{BH S-matrix and QNM-pole splitting method}
\label{subsec:pole-splitting}
%%%%%%%%%%%%%%%%%%%%%%%%%%%%

Let us consider the spectrum $Z_{\ell m \omega}$ in Eq.~\eqref{spectrum_original_for_far}:
\begin{equation}
    Z_{\ell m \omega} = \hat{D}_{\ell m \omega} {\cal T}_{\ell m \omega}\,,
    \label{ZDT}
\end{equation}
where
\begin{align}
    \hat{D}_{\ell m \omega} &= \frac{B_{\ell m \omega}^{\rm out}}{2 i \omega B_{\ell m \omega}^{\rm in}}\,,
    \label{eq:Ddef}\\
    {\cal T}_{\ell m \omega} &= \int_{r_+}^{\infty} dr' \frac{R_{\ell m \omega}^{\rm in}}{B_{\ell m \omega}^{\rm out}} \Delta^{-2} T_{\ell m \omega}\,.
    \label{eq:Tdef}
\end{align}
The first factor is the BH response and the second is the source.
This is the pair that Sec.~\ref{sec:Pole-Splitting-Method} splits.
In the following, we omit the subscripts $\ell m \omega$ for simplicity.
The factor $\hat{D}$ can be split into pole and non-pole sectors as
\begin{equation}
\label{eq:Dsplit}
\hat{D} = \underbrace{\sum_{n} \frac{E_{n}}{(\omega - \omega_{n})}}_{\text{pole sector: } \hat{D}_{\rm pole}} + \underbrace{\left[ \hat{D} - \sum_{n} \frac{E_{n}}{(\omega - \omega_{n})} \right]}_{\text{non-pole sector: }\hat{D}_{\rm non}}\,,
\end{equation}
where $E_n$ is the QNM excitation factor defined as the residue of $\hat{D}$ at QNM poles:
\begin{equation}
E_n \coloneqq \left. \frac{B_{\ell m \omega}^{\rm out}}{ 2i \omega (d B_{\ell m \omega}^{\rm in}/ d \omega)} \right|_{\omega = \omega_n}\,.
\label{QNMEFs_definition}
\end{equation}
Then we split the waveform $\Psi_{4}(u)$ as
\begin{align}
    &\Psi_{4}(u) \sim \int d \omega e^{-i \omega u} Z (\omega)\,,\\
    &= \underbrace{\int d\omega e^{-i \omega u} \hat{D}_{\rm pole} (\omega) {\cal T} (\omega)}_{\eqqcolon \Psi_{\rm pole} (u)} 
    + \underbrace{\int d\omega e^{-i \omega u} \hat{D}_{\rm non} (\omega) {\cal T} (\omega)}_{\eqqcolon \Psi_{\rm non} (u)}\,.
\end{align}
All the QNM poles are contained in $\Psi_{\rm pole}$, whereas $\Psi_{\rm non}$ contains the remaining non-pole contributions. Since this decomposition is defined in the frequency domain, it should not be interpreted as a decomposition into late- and early-time signals. The physical interpretation of each sector is discussed in Sec.~\ref{subsec:physical-meaning}.

We note that the decomposition introduced in Ref.~\cite{Su:2026gmp} is different from the pole-splitting method employed here.
Their decomposition is based on the causal structure of the Green's function.
For a plunge source, the full real-axis inverse Fourier transform is decomposed into the direct and QNM-plus-tail sectors according to the causal support associated with each source location.
On the other hand, our pole-splitting method explicitly isolates QNM poles in the BH S-matrix, factored out from the full spectrum.
%

%%%%%%%%%%%%%%%%%%%%%%%%%%%%
\subsection{Physical meaning of the pole-splitting method}
\label{subsec:physical-meaning}
%%%%%%%%%%%%%%%%%%%%%%%%%%%%

We now discuss the physical interpretation of the two sectors, $\Psi_{\rm pole}$ and $\Psi_{\rm non}$.
Their physical contents at early and late times are summarized in Table~\ref{tab:sector_summary}.

Equations~\eqref{eq:psi-pole-sec}-\eqref{eq:psi-non-sec} are the single-segment counterparts of the waveform sectors $\Psi_{\rm pole}$ and $\Psi_{\rm non}$ of Sec.~\ref{subsec:pole-splitting}, and Eq.~\eqref{eq:DW-shares} says that both have a share of the DW.
What distinguishes the two sectors is therefore not the presence of the DW but the terms oscillating at the QNM frequencies, which Eq.~\eqref{eq:psi-pole-sec} assigns to $\Psi_{\rm pole}$ alone.
The splitting removes the QNM components from $\Psi_{\rm non}$, so the instantaneous frequency of $\Psi_{\rm non}$ is expected to reflect that of the DW, whereas that of $\Psi_{\rm pole}$ is contaminated by QNMs.

At late times, $\Psi_{\rm pole}$ describes the conventional ringdown, consisting of the QNMs together with the direct-wave component originating from the QNM-excitation region or the near-horizon region.
The remaining sector, $\Psi_{\rm non}$, contains the direct-wave component together with the branch-cut contribution responsible for the
power-law tail.

\begin{figure}[t]
\centering
\includegraphics[width=0.95\linewidth]{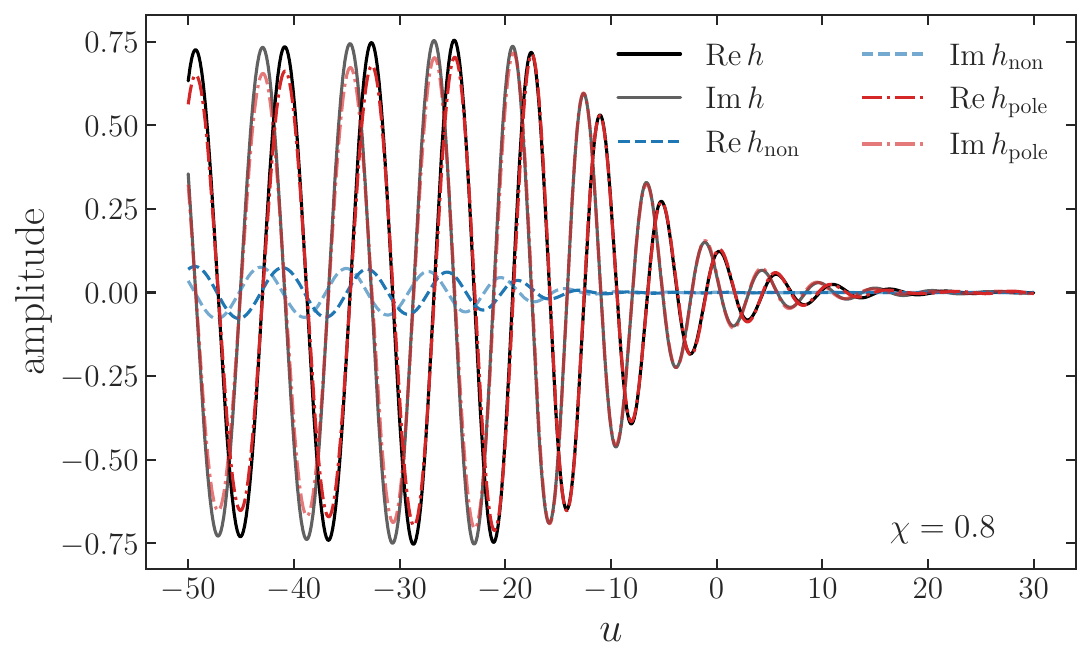}
\caption{
Split waveforms, $h_{\rm non} (u)$ and $h_{\rm pole} (u)$, and the full waveform $h (u)$ sourced by the extreme-mass-ratio merger with the massive BH ($\chi = 0.8$). The angular mode is set to $\ell = m =2$. 
}
\label{fig_wave_split_0800}
\end{figure}

At early times, the interpretation is different.
Although the waveform is not yet dominated by ringdown, $\Psi_{\rm pole}$ generally remains nonzero.
In particular, for continuously driven systems such as EMRIs, gravitational waves emitted by the orbiting particle are scattered by the BH geometry, exciting QNMs well before the plunge.
These pre-plunge QNM excitations are therefore also contained in $\Psi_{\rm pole}$.
In contrast, $\Psi_{\rm non}$ contains only the prompt/direct-wave component and does not include any QNM contribution.
When the source contains little low-frequency power, the branch-cut contribution is also strongly suppressed, so that $\Psi_{\rm non}$ is dominated by the DW.
The pole-splitting method has been applied to the GW waveform generated by a particle undergoing a quasicircular plunge in the extreme-mass-ratio regime (see FIG.~\ref{fig_wave_split_0800}).

This property makes the pole-splitting method particularly useful.
Since it introduces no artificial and frequency-dependent time shift, inherent in the QNM-filtering method, the instantaneous frequency extracted from $\Psi_{\rm non}$ directly probes the dynamical evolution of the DW, as demonstrated in Sec.~\ref{sec:Numerical-direct-wave}.
In contrast, the frequency evolution extracted from $\Psi_{\rm pole}$ is generally contaminated by QNM contributions even before the onset of the late-time ringdown.

%%%%%%%%%%%%%%%%%%%%%%%%%%%%
\section{Numerical test of DWs via the pole-splitting method}
\label{sec:Numerical-direct-wave}
%%%%%%%%%%%%%%%%%%%%%%%%%%%%

\begin{table*}[t]
\centering
\caption{Contents of $h_{\rm pole}$ and $h_{\rm non}$ at early and late times.}
\begin{tabular}{cll}
\toprule
Sector & Early times (pre-plunge) & Late times (post-plunge)\\
\midrule
$h_{\rm pole}$ &
Direct wave + pre-plunge QNMs &
Direct wave + ringdown QNMs\\

$h_{\rm non}$ &
Direct wave &
Direct wave + tail\\
\bottomrule
\end{tabular}
\label{tab:sector_summary}
\end{table*}

We numerically analyze the DW signals sourced by a quasi-circular-plunging orbit in the Kerr spacetime.
Our analysis does not rely on the rational/full QNM filtering, which causes a non-trivial time shift in the resulting waveforms.
Instead, we utilize a novel technique to split the whole waveform into two sectors: non-pole waveform $\Psi_{\rm non} (u)$ and pole waveform $\Psi_{\rm pole} (u)$ defined in Sec.~\ref{subsec:pole-splitting}.

\subsection{DWs from an extreme-mass-ratio merger in Kerr background}
We compute the GW waveforms in strain amplitude $h(u)$, whose spectrum, $\tilde{h}$, is given by 
\begin{equation}
    \tilde{h} = -\frac{2}{r} \frac{Z_{\omega}}{\omega^2} \,,
    \label{strain_definition}
\end{equation}
and we utilize the source term of the quasi-circular orbit obtained by Watarai in Ref.~\cite{Watarai:2024huy} based on the Ori-Thorne method \cite{Ori:2000zn}.
Here, we also redefine the function $\hat{D}$ in Eq.~\eqref{eq:Ddef} by absorbing the factor of $1/\omega^2$ appearing in the strain~\eqref{strain_definition}, i.e., $\hat{D}/\omega^2 \to \hat{D}$.
The source term is provided in the Sasaki-Nakamura formalism \cite{Sasaki:1981sx}, but it can be transformed into the Teukolsky variable after the whole computation.
To obtain the transfer functions, $\hat{D}_{\rm pole}$ and $\hat{D}_{\rm non}$, we compute the QNM excitation factors, $E_{n}$ in Eq.~\eqref{QNMEFs_definition}, with the MST method \cite{Mano:1996vt}.
To construct the transfer functions, we truncate the overtones at $n = 20$ for prograde modes and at $n=10$ for retrograde modes (for the convergence of the pole expansion, see Appendix~\ref{app_detail_PSmethod}).
We then decompose the full strain into the pole and non-pole sectors, $h_{\rm pole}(u)$ and $h_{\rm non}$, respectively.

\begin{figure}[t]
\centering
\includegraphics[width=1\linewidth]{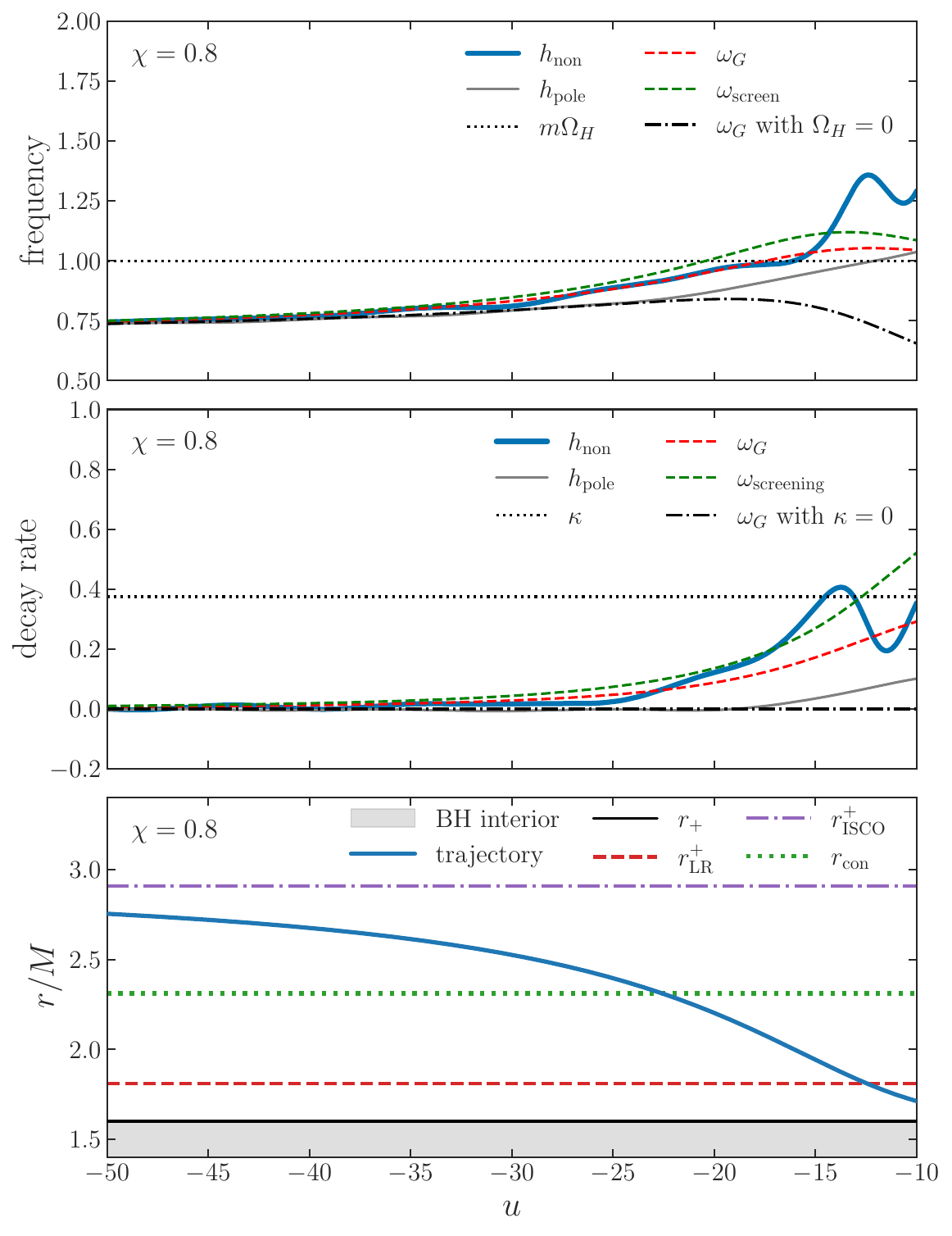}
\caption{
\textbf{Top:} Evolution of the frequency of $h_{\rm non}$ (blue thick solid) and $h_{\rm pole}$ (gray solid).
For comparison, $\mathrm{Re}[\omega_{\rm G}(u)]$ (red dashed), $\mathrm{Re}[\omega_{\rm screen}(u)]$ (green dashed), and $\mathrm{Re}[\omega_{\rm G}(u)]$ with $\Omega_H=0$ (black dash-dotted) are also shown.
The horizontal dotted line indicates $m\Omega_H$.
\textbf{Middle:} Evolution of the decay rates of $h_{\rm non}$ and $h_{\rm pole}$, together with the corresponding predictions from $\omega_{\rm G}(u)$, $\omega_{\rm screen}(u)$, and $\omega_{\rm G}(u)$ with $\kappa=0$.
The horizontal dotted line indicates $\kappa$.
\textbf{Bottom:} Radial trajectory of the ISCO-plunge particle (blue solid).
The radii of the prograde light ring ($r_{\rm LR}^{+}$), ISCO ($r_{\rm ISCO}^{+}$), and outer horizon ($r_{+}$) are indicated.
The QNM convergence radius $r_{\rm con}$ (green dotted) is also shown 
as a reference for the effective location of the potential barrier 
relevant to QNM excitation.\footnote{
The values of $r_{\rm con}$ evaluated by one of the authors are tabulated in Table~I in Ref.~\cite{Oshita:2026vxh}.}
Here we set $\ell=m=2$ and the dimensionless spin of the massive BH to $\chi=0.8$.
}
\label{fig_DW_1_spins}
\end{figure}

The result of the pole-splitting analysis is shown in FIG.~\ref{fig_wave_split_0800}.
We also compute the time evolution of the frequency and decay rate of the individual sectors in FIG.~\ref{fig_DW_1_spins}.
The dynamical complex frequency for DWs, $\omega_{\rm G} (u)$ [see Eq.~\eqref{definition_omega_G}], for the plunging orbit is also shown as a reference (dashed line in FIG.~\ref{fig_DW_1_spins}).
We also incorporate the zero of the screening factor $P(\omega)$ [see Eq.~\eqref{eq:Wpoly}] at the complex horizon frequency, $\omega = \omega_{\rm H}^{(1)}$ (for the details of the screening effect, see Appendix~\ref{app_screening_SN}).
In the near-horizon regime, the strain can be approximated as [see Eqs.~\eqref{mm_components}, \eqref{spectrum_final_form} and \eqref{psi_4_dw_approx}]
\begin{align}
\begin{split}
h \sim h_{\rm screen} \coloneqq  (\omega_{\rm G} - \omega_{\rm H}^{(1)}) e^{-i \int \omega_{\rm G} du}\,.
\end{split}
\label{screening_factor_approx}
\end{align}
where the zero of $P(\omega)$ at $\omega = \omega_{\rm H}^{(1)}$ is approximated by the factor $\omega_{\rm G} - \omega_{\rm H}^{(1)}$.
It affects the frequency and decay rate of DWs, and their time evolution can be evaluated by the real and imaginary parts of $\omega_{\rm screen}$ defined as
\begin{equation}
    \omega_{\rm screen} (u) \coloneqq i \frac{d}{du} \log (h_{\rm screen})\,.
\end{equation}

\begin{figure*}[t]
\centering
\includegraphics[width=0.49\linewidth]{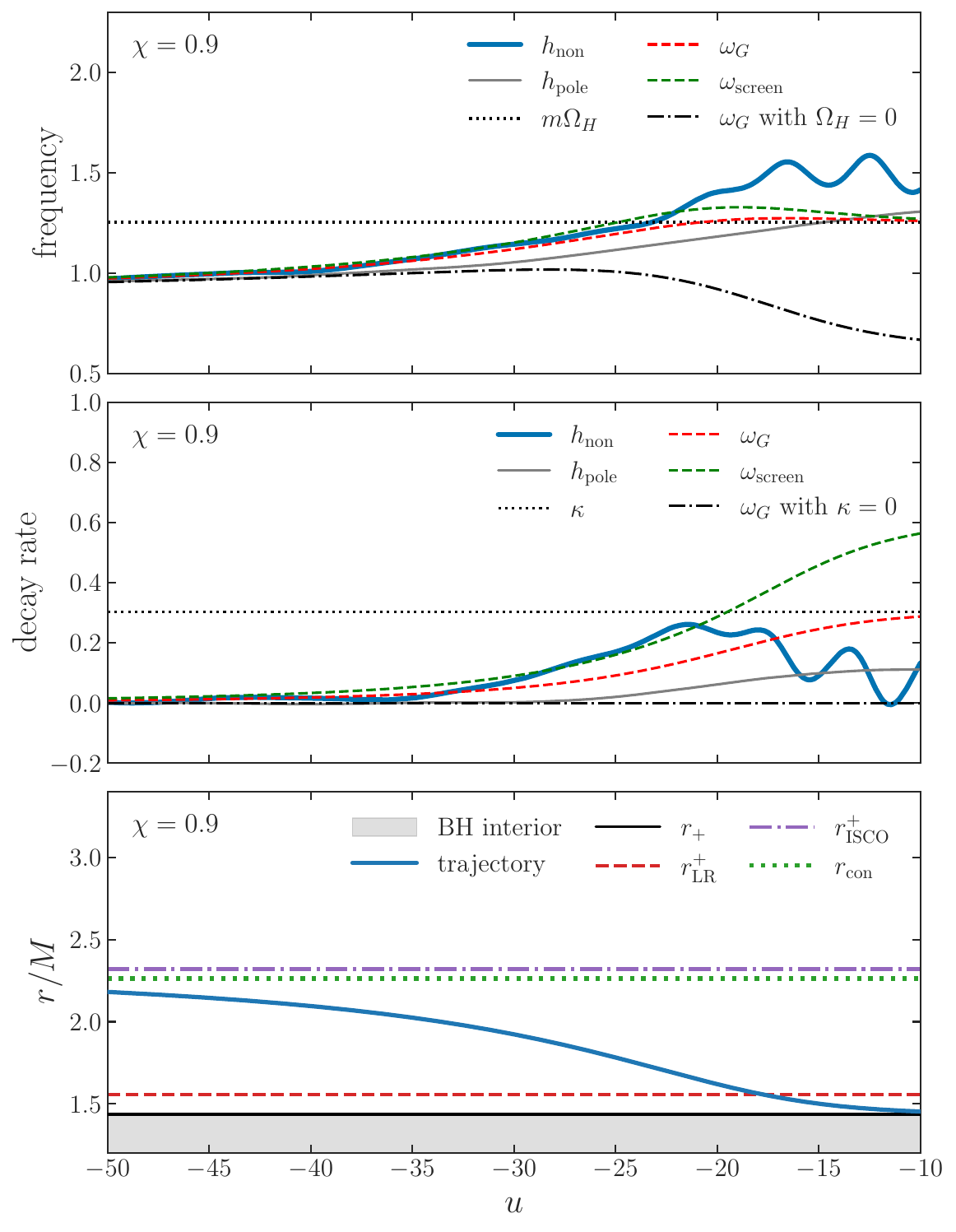}
\includegraphics[width=0.482\linewidth]{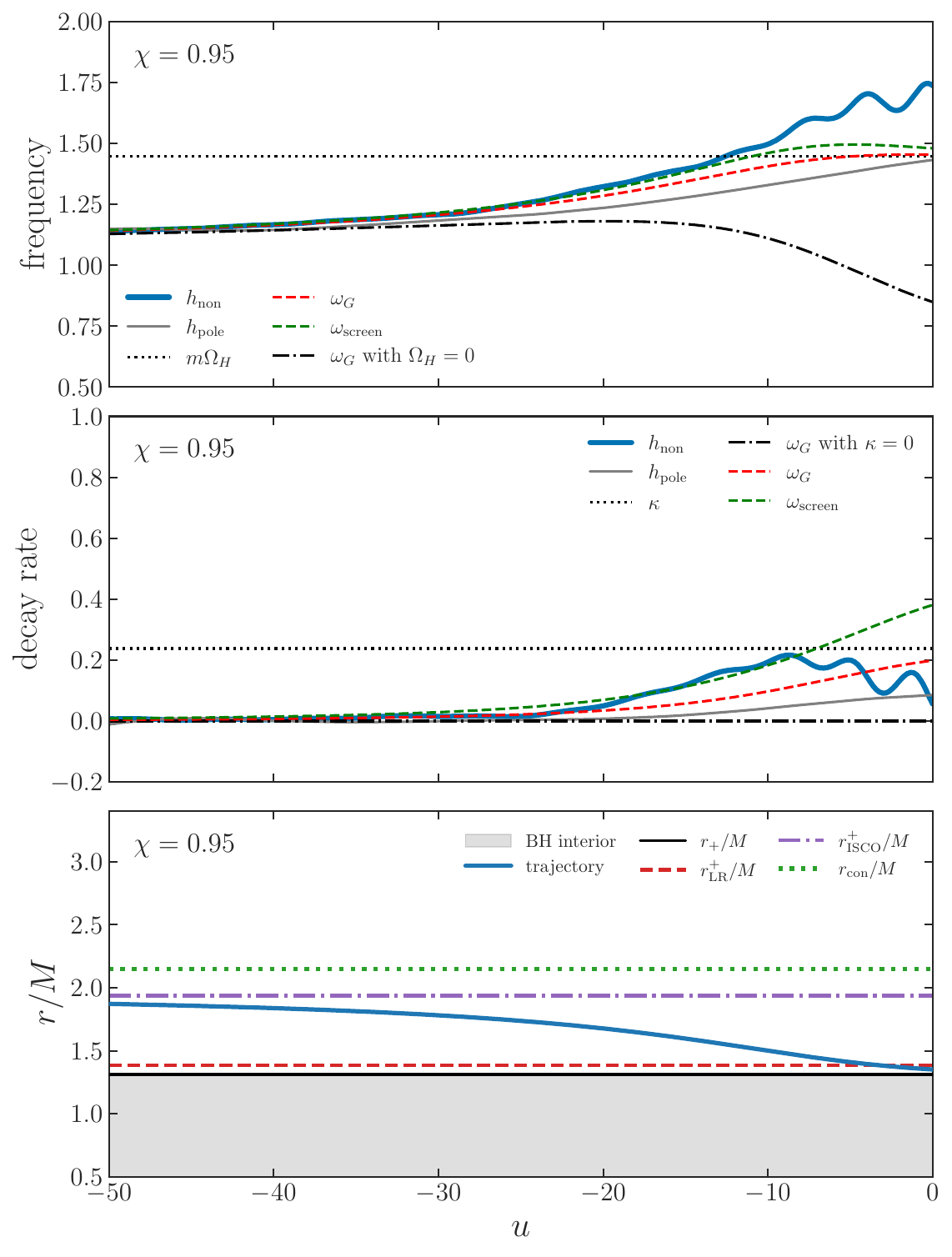}
\caption{
Same as FIG.~\ref{fig_DW_1_spins}, but for higher BH spins:
$\chi=0.9$ (left) and $\chi=0.95$ (right).
The other parameters are the same as in FIG.~\ref{fig_DW_1_spins}.
For $\chi=0.95$, the value $r_{\rm con}= 2.147M$ is obtained by 
interpolating the values reported in Table~I of Ref.~\cite{Oshita:2026vxh}  using \texttt{numpy.interp}.
}
\label{fig_DW_2_spins}
\end{figure*}
It is confirmed that the time evolution of the direct-wave frequency, $\omega_{\rm G} (u)$, agrees with that of $h_{\rm non}$ for $\chi =0.8$.
On the other hand, for $\chi = 0.9$ and $0.95$, the time evolution of $\omega_{\rm screen}$ becomes more consistent with that of $h_{\rm non}$ as shown in FIG.~\ref{fig_DW_2_spins}.
This difference can be understood from the radial location of the particle relative to the effective potential barrier.
Here, the location of the potential barrier relevant to QNM excitation should not necessarily be identified with the prograde light-ring radius $r_{\rm LR}^{+}$ for a finite $\ell$ \cite{Oshita:2026vxh}.
For this reason, we show the Kerr-QNM convergence radius $r_{\rm con}$, introduced and evaluated in Ref.~\cite{Oshita:2026vxh}, in the bottom panels as a reference for the effective location of the potential barrier relevant to QNM excitation.
For $\chi=0.8$, the particle is located outside the QNM convergence region, $r=r_{\rm con}$, during a large part of the time interval.
For $\chi=0.9$ and $0.95$, on the other hand, the particle is already inside $r_{\rm con}$ over most of the corresponding time intervals.
This provides a natural explanation for the difference of the case in $\chi = 0.8$ and in $\chi \gtrsim 0.9$.
When the GW source is outside the effective potential barrier, the outgoing DW can propagate directly toward the observer without being significantly filtered by the barrier, and hence $\omega_{\rm G}$ provides a good description of its frequency evolution.
Once the source moves inside the barrier, however, the outgoing DW has to propagate through the barrier before reaching the observer.
Its frequency evolution is then affected by the frequency-dependent greybody factor, making $\omega_{\rm screen}$ a better description of the observed DW.
The screening effect occurs when the particle or GW source passes the potential barrier of the radial perturbation equation, located around the light ring, and when the frequency of DWs approaches the horizon mode.
We find that the time evolution of the frequency of $h_{\rm pole}$ (grey solid in FIGs.~\ref{fig_DW_1_spins} and \ref{fig_DW_2_spins}) differs from that of $h_{\rm non}$ and does not match $\omega_{\rm G}$.
This observation supports the idea that our pole-splitting analysis works to split between the DW component and others.
The growth of the frequency is consistent with the frame-dragging effect in the ergoregion, which is associated with the frequency $ \Omega_{\rm H}$.
Our result suggests that the ergoregion in Kerr spacetime can be probed via the frequency evolution in $h_{\rm non}$ radiated by the extreme-mass-ratio merger.

The agreement in the real part of the instantaneous frequency demonstrates that the particle-based picture successfully captures the dominant dynamics of the DW. 
On the other hand, the agreement in the imaginary part for $\chi = 0.8$ remains only qualitative (FIG.~\ref{fig_DW_1_spins}), suggesting that a more complete treatment of wave-propagation effects is required for a quantitative description.
One possible origin of the remaining discrepancy is the approximate treatment of $\hat{D}$ or the black-hole S-matrix.
We provide the details of the screening effect in Appendix~\ref{app_screening_SN} and discuss the origin of the factor in the Teukolsky and Sasaki-Nakamura formalism.
The approximation Eq.~\eqref{screening_factor_approx} captures the screening effect of the BH geometry against the horizon mode, but it does not fully incorporate the next-to-leading-order effect, namely the full frequency-dependence of the Green's function.
In the near-horizon case, the DW in the non-pole component can be approximated as 
\begin{equation}
h_{\rm non}(u)
\sim
C \left[\omega_{\rm G}(u)\right] [\omega_{\rm G}(u) - \omega_{\rm H}^{(1)}]
\exp\left[-i\int^u\omega_G(u') du'\right]\,,
\end{equation}
where $C(\omega)$ corresponds to the frequency-dependent factor in Eq.~\eqref{psi_4_dw_approx} that we neglect in Eq.~\eqref{screening_factor_approx}.
The instantaneous complex frequency extracted by the pole-splitting method is
\begin{align}
\begin{split}
\omega_{\rm non}(u) = i\frac{\dot{h}_{\rm non}}{h_{\rm non}}
=\omega_G(u)
&+
i\frac{d}{du} \ln  \left[\omega_{\rm G}(u) - \omega_{\rm H}^{(1)}\right]\\
&+
i\frac{d}{du} \ln C \left[\omega_{\rm G}(u)\right]\,,
\end{split}\label{ome_non}
\end{align}
and the second term reduces to
\begin{equation} i\frac{d}{du} \ln  \left[\omega_{\rm G} - \omega_{\rm H}^{(1)} \right]
\simeq -i \kappa\,.
\label{approx_hat_D}
\end{equation}
Thus, the frequency dependence of $C(\omega)$ also modifies both the real and imaginary parts of the extracted frequency. 
The variation of its magnitude contributes directly to the imaginary part, while the variation of its phase modifies the real part. 
Here we take into account only the first and second terms in Eq.~\eqref{ome_non}, and hence the incomplete modeling of these variations may account, at least partly, for the less accurate agreement found in the imaginary part.
A noticeable deviation in the real part appears once $\rm{Re}(\omega_{\rm G})$ approaches the superradiant threshold $m\Omega_H [=\operatorname{Re}(\omega_H)]$.
This behavior can be understood as numerical errors in the pole-splitting method or waveform computation, or as a consequence of the nontrivial evolution of the phase of the black-hole S-matrix, which becomes non-negligible in this frequency regime but is not fully incorporated in the present model.
Although the present approximation incorporates the zero of $\widetilde{\mathcal{W}}_i$ at $\omega=\omega_{\rm H}^{(1)}$, it may not be sufficient to describe the rapid variation of the phase or amplitude of $\hat{D}(\omega)$ around the superradiant threshold. 
Incorporating its full frequency dependence may also require special care, as it depends on the perturbation variables describing the BH S-matrix, e.g., the Teukolsky, Sasaki-Nakamura, and Chandrasekhar-Detweiler variables.
A quantitative analysis based on the full BH S-matrix is left for future work.

Nevertheless, it is remarkable that even this leading-order approximation successfully reproduces the time evolution of the real and imaginary parts of the source-driven frequency. 
Such a comparison is enabled by the pole-splitting method, which does not introduce a non-trivial time-shift effect or a non-trivial deformation of the original waveform inherent in the QNM filtering method.
This suggests that the time evolution of the DW may provide a direct probe of the frame-dragging effect in the ergoregion, as characterized by the horizon frequency $\Omega_{\rm H}$.

%%%%%%%%%%%%%%%%%%%%%%%%%%%%
\section{Conclusion}
\label{sec:Conclusion}
%%%%%%%%%%%%%%%%%%%%%%%%%%%%
We have studied direct waves (DWs) emitted by a particle plunging into a Kerr black hole in the linear, extreme-mass-ratio regime.
We have aimed to organize the theoretical description of DW emission, discuss its relation to dynamical QNM excitation, and examine how to identify its frequency evolution in the waveform.
We addressed these questions by analyzing the near-horizon spectrum and numerically calculating the waveforms of quasi-circular plunges.

The near-horizon description characterizes the DW by the source-driven frequency $\omega_{\rm G}$~\cite{Oshita:2025qmn}.
The real part of $\omega_{\rm G}$ reflects the angular motion and frame dragging, while its imaginary part describes the decay associated with radial infall and gravitational redshift.
As the particle approaches the horizon, $\omega_{\rm G}$ tends to $\omega_{\rm H}^{(1)}=m\Omega_{\rm H}-i\kappa$.
The near-horizon source weight contains a screening factor that vanishes at this frequency.

We discussed the relation between DWs and dynamical QNM excitation by comparing the frequency-domain Green's function method and the time-domain Green's function method.
The frequency-domain method identifies the DW term and its amplitude, while the time-domain method expresses the QNM response in terms of time-dependent amplitudes.
Introducing a pole/non-pole decomposition into the frequency-domain result separates the DW amplitude into two contributions.
The pole-sector contribution coincides with the term oscillating at the source-driven frequency in the time-domain QNM response.
This contribution is therefore part of dynamical QNM excitation~\cite{DeAmicis:2025xuh}.

The decomposition also identifies the non-pole sector as a candidate for extracting the DW frequency evolution.
The pole response contains terms at both $\omega_{{\rm G},i}$ and the QNM frequencies.
The non-pole response contains the exponential at $\omega_{{\rm G},i}$ and the cut and arc integrals.
Neglecting the integral contributions to the waveform and its time derivative leaves the term oscillating at the source-driven frequency.
Under this approximation, and when the time dependence of the remaining amplitude can be neglected, the instantaneous frequency of the non-pole waveform is expected to trace the DW frequency evolution.

With the complete point-particle source, we have extended the horizon-mode pole cancellation of Kubota {\it et al.}~\cite{Kubota:2026qjh} to all radial and source orders under the conditions specified in Sec.~\ref{sec:NH-cancellation}.
Although higher-order radial coefficients can themselves have poles at the horizon-mode frequencies, the differential structure of the complete source preserves the cancellation.
The DW model instead combines leading near-horizon radial and source expressions with a first-order time expansion of $\phi(t)$ and $x(t)$ along the plunge, without requiring a convergent all-order horizon expansion of the orbit.
We have not quantified the domain of validity of these combined approximations or their relation to the convergence domain of the horizon expansion, and assess the frequency model over finite intervals by numerical comparison.

We introduced the pole-splitting method to separate waveform contributions according to the analytic structure of the transfer function.
The QNM poles and their residues determine the pole terms, and subtraction of these terms defines the remainder.
Applying this decomposition to the waveform spectrum gives pole and non-pole waveforms on the same retarded-time coordinate.
The method subtracts the contributions associated with the selected QNM poles without fitting mode amplitudes in the time domain.

We tested the method on the $\ell=m=2$ strain from quasi-circular plunges with $\chi=0.8$, $0.9$, and $0.95$.
At $\chi=0.8$, the real part of the non-pole instantaneous frequency follows $\operatorname{Re}\omega_{\rm G}$ over part of the analyzed interval, before the late-time deviations.
At $\chi=0.9$ and $0.95$, the screened prediction $\omega_{\rm screen}$ provides a closer description over part of the analyzed interval.
This difference is consistent with the source locations relative to the QNM convergence radius~\cite{Oshita:2026vxh} and supports the interpretation of curvature-barrier screening discussed in Sec.~\ref{sec:Numerical-direct-wave}.
The pole waveform shows a different frequency evolution.
This behavior is consistent with interference between the terms at the source-driven frequency and the QNM frequencies.
The qualitative agreement of the non-pole frequency with the DW predictions supports its potential use as a probe of frame dragging and gravitational redshift near the horizon.

The present DW model does not reproduce the numerical results quantitatively throughout the analyzed interval.
The decay rate at $\chi=0.8$ agrees only qualitatively, and deviations become apparent as the real part of $\omega_{\rm G}$ approaches $m\Omega_{\rm H}$.
The analytic prediction omits the frequency dependence of the black-hole response beyond the screening zero and neglects the branch cut and large-arc contributions.
Numerical errors in the waveform calculation and pole splitting may also affect the extracted frequency.
We leave a quantitative explanation of these deviations to future work.

The use of pole splitting is not restricted to DW analysis.
Its ability to subtract linear QNM contributions without fitting their amplitudes could provide a way to examine the remaining radiation in black-hole ringdowns.
%

%%%%%%%%%%%%%%%%%%%%%%%%%%%%%%%%%%%%%%%%%%%%%%%%%%%%%%%%%
\section{Acknowledgments}
%%%%%%%%%%%%%%%%%%%%%%%%%%%%%%%%%%%%%%%%%%%%%%%%%%%%%%%%%
%
N.~N. was supported by the Create the Future Project at Kyoto University.
N.~O. was supported by Japan Society for the Promotion of Science (JSPS) KAKENHI Grant No.~JP26K17142.
H.~T. is supported by the Hakubi project at Kyoto University, and by JSPS KAKENHI Grant No.~JP26K17146.

We used \texttt{ChatGPT GPT-6 Astra} and \texttt{ChatGPT GPT-5.6 Sol} to assist with literature review, coding, and the analytical calculations in Sec.~\ref{sec:NH-cancellation}, all closely supervised by the authors.

\appendix

%%%%%%%%%%%%%%%%%%%%%%%%%%%%
\section{Details of the pole-splitting method}
\label{app_detail_PSmethod}
%%%%%%%%%%%%%%%%%%%%%%%%%%%%
%
We describe the numerical prescription for separating the waveform into the pole and non-pole sectors.
The main difficulty is that the numerical errors at high frequencies in the (effective) source factor $Z(\omega) / \hat{D}(\omega)$ are strongly amplified.
We therefore determine the upper integration frequency from a waveform-reconstruction criterion before performing the pole splitting.

For a fixed spheroidal mode, $(\ell,m)$, we can write the full spectrum as
\begin{equation}
    Z(\omega) = \hat{D} (\omega) {\cal T} (\omega)\,,
\end{equation}
where ${\cal T} (\omega)$ is a source part that depends on the trajectory of a particle [see also Eq.~\eqref{ZDT}].
In the numerical implementation, we compute
\begin{equation}
    {\cal T} (\omega) = \frac{Z(\omega)}{\hat{D} (\omega)}\,,
    \label{app_T}
\end{equation}
and obtain
\begin{align}
    Z_{\rm pole} (\omega) &= {\cal T} (\omega) \hat{D}_{\rm pole} (\omega)\,,\\
    Z_{\rm non} (\omega) &= {\cal T} (\omega)  \left[ \hat{D} (\omega) - \hat{D}_{\rm pole} \right]\,.
\end{align}
\begin{figure*}[t]
\centering
\includegraphics[width=0.9\linewidth]{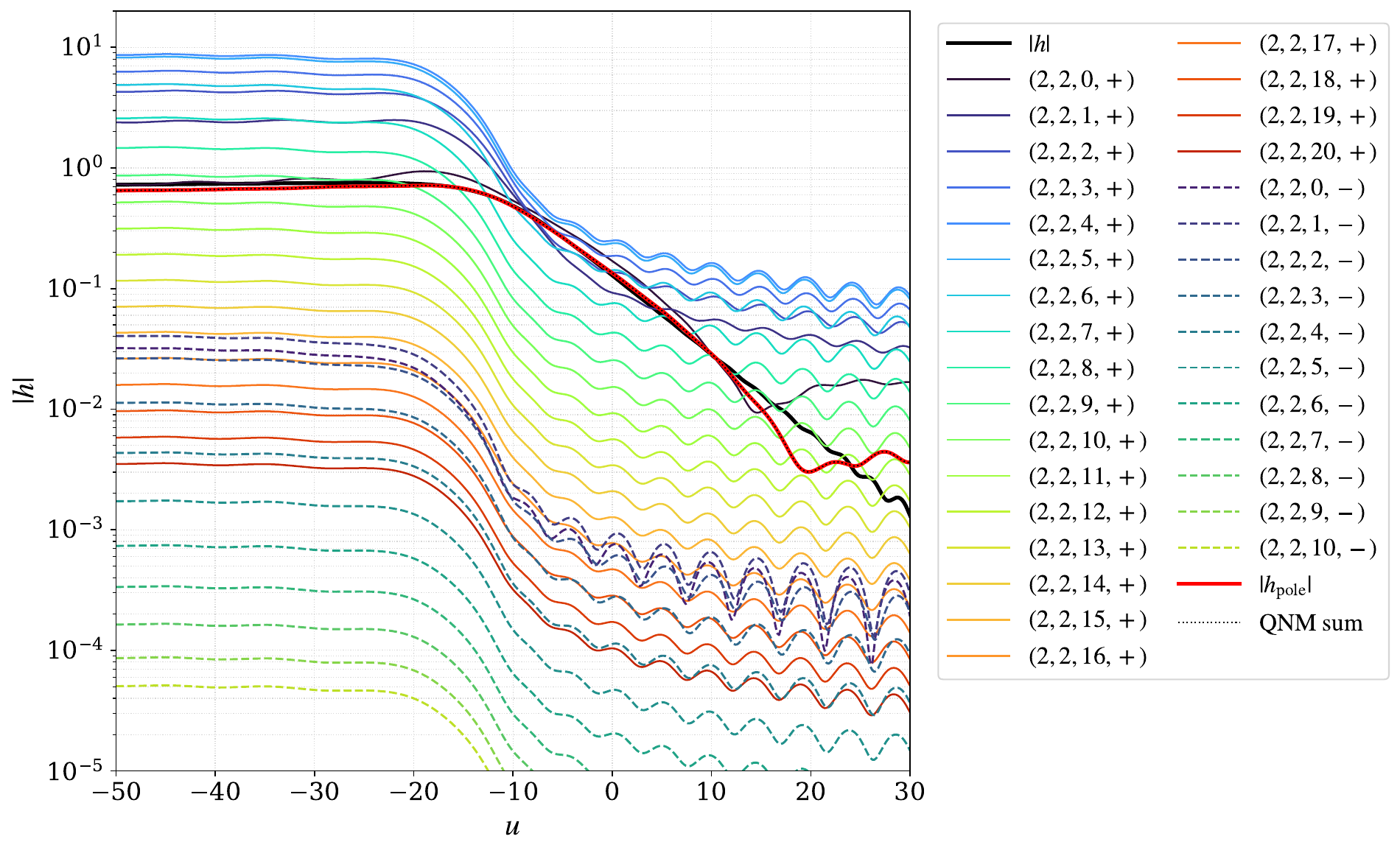}
\caption{
Convergence of the pole (QNM) contribution in $\Psi_{\rm pole}$ for $\chi =0.8$ and $\ell=m=2$. Each QNM contribution is labeled by $(\ell,m,n,\pm)$, where $+$ and $-$ represent the prograde and retrograde QNM, respectively.
}
\label{fig_app_qnm_decomp}
\end{figure*}
In Eq.~\eqref{app_T}, the numerical errors of $Z(\omega)$ are exponentially enhanced at high frequencies due to the factor of $1/\hat{D}(\omega)$.
Extending the Fourier mode to the largest available value can therefore contaminate the separated waveforms.
To suppress the enhanced contamination, we set the upper cutoff by requiring a truncated spectrum to reconstruct the original waveform with a preferred accuracy.
The original waveform is calculated over the broad frequency interval $[\omega_{\rm min}, \omega_{\rm max}]$.
\begin{equation}
    \Psi_{\rm original} = \int_{\omega_{\rm min}}^{\omega_{\rm max}} d \omega Z(\omega) e^{-i \omega u}\,.
\end{equation}
For a prograde plunge, we consider in this work that the trial waveform is evaluated with $\omega_{\rm min} = 0$, as the retrograde part is negligible:
\begin{equation}
    \Psi_{\rm tr} = \int_0^{\omega_{\rm tr}} d \omega Z(\omega) e^{-i \omega u}\,.
\end{equation}
To obtain a minimum value of $\omega_{\rm tr}$, we quantify the reconstruction error by the mismatch ${\cal M}$ defined by
\begin{equation}
    {\cal M} (\omega_{\rm tr}) = \left| 1- \frac{\braket{\Psi_{\rm original} , \Psi_{\rm tr}}}{\sqrt{\braket{\Psi_{\rm original} , \Psi_{\rm original}} \braket{\Psi_{\rm tr} , \Psi_{\rm tr}}}} \right|\,,
\end{equation}
where 
\begin{equation}
    \braket{f,g} = \int_{u_{\rm min}}^{u_{\rm max}} d u f(u) g^*(u)\,.
\end{equation}
We start from $\omega_{\rm tr} = \text{Re} (\omega_{\ell m 0})$ and increase the cutoff in steps of $\Delta \omega = 0.01$.
It stops at the first value satisfying
\begin{equation}
    {\cal M} (\omega_{\rm tr}) < {\cal M}_{\rm tol}\,,
\end{equation}
with ${\cal M}_{\rm tol} = 10^{-4}$.
For the representative case $a = 0.8$ and $(\ell,m) = (2,2)$, using $[\omega_{\rm min}, \omega_{\rm max}] = [-1,2]$ and $[u_{\rm min}, u_{\rm max}] = [-50,30]$, we have the optimal value $2M\omega_{\rm tr} = 1.46$.
We then compute the pole and non-pole sectors in strain amplitude, denoted by $h_{\rm pole}$ and $h_{\rm non}$, respectively.

FIG.~\ref{fig_app_qnm_decomp} demonstrates the convergence of the pole-splitting method.
The prograde and retrograde QNMs, up to $n=20$ and $n=10$, respectively, are included in the pole-splitting analysis.
These numbers are large enough to see the convergence of the waveform reconstruction (see also the previous work in Ref.~\cite{Oshita:2024wgt}).
We see that the late-time frequency and decay rate of an individual QNM contribution do not generally coincide with $\omega_{\ell mn}^{\pm}$ although the fundamental-mode component follows the decay rate of ringdown.
This is because the inverse Fourier integral is restricted to the finite frequency interval, $[0, \omega_{\rm tr}]$, and an individual QNM contribution contains end-point-induced terms.
Despite the non-QNM term, their coherent sum accurately reconstructs the ringdown waveform as shown in FIGs.~\ref{fig_wave_split_0800} and \ref{fig_app_qnm_decomp}.
This is enabled by cancellations among the end-point-induced terms, which originate from correlations among the QNM excitation factors (see FIG.~9 in Ref.~\cite{Oshita:2024wgt}).

\section{Screening of the horizon mode}
\label{app_screening_SN}

The screening factor of the horizon mode appears via the source term in the Teukolsky formalism, as was pointed out by Zimmerman {\it et al.}~\cite{Zimmerman:2011dx} and was reviewed in Sec.~\ref{sec_Theory}.
On the other hand, the same screening factor appears from the BH greybody factor.
To see this, let $B^{\rm in}_{\rm SN}$ and $B^{\rm in}$ denote the incoming-wave amplitudes of the homogeneous solutions that are ingoing at the horizon for the Sasaki–Nakamura and Teukolsky variables, respectively.
Their asymptotic transformation is given by 
\begin{align}
\label{eq:Bin-transform}
    \frac{1}{B^{\rm in}_{\rm SN}} = \frac{d(\omega)}{B^{\rm in}}\,,
\end{align}
where
\begin{align}
 d &\coloneqq \frac{(2Mr_+)^{5/2}}{\omega^2}
 \left[
 k_{\rm H}^2+6i\epsilon_{\rm H} k_{\rm H}-8\epsilon_{\rm H}^2
 \right]\,,\\
 \epsilon_{\rm H} &\coloneqq \frac{\sqrt{M^2 - a^2}}{4Mr_+}\,.
\end{align}
Since the surface gravity $\kappa$  is represented as $\kappa = 2 \epsilon_{\rm H}$, this coefficient can be written as
\begin{equation}
    d = \frac{(2Mr_+)^{5/2}}{\omega^2}
 (k_{\rm H}+i\kappa)(k_{\rm H}+2i\kappa)\,.
\end{equation}
Thus, precisely the same screening factor, $(k_{\rm H}+i\kappa)(k_{\rm H}+2i\kappa)$, that appears in the near-horizon source term in the Teukolsky formalism, also appears in the greybody factor.

In the Teukolsky formalism, it is identified in the near-horizon source term, whereas in the Sasaki-Nakamura formalism, it is already contained in the greybody factor.
Of course, the full observable waveform, which is determined by the convolution of the source term and the Green's function, is independent of this formalism-dependent separation.

The screening factor appears in the near-horizon approximation of the source in the Teukolsky formalism, while the Sasaki-Nakamura formalism incorporates it into the greybody factor.
We therefore attribute the screening to the BH geometry, or to a curvature potential barrier around the light ring.

In the MST method of Ref.~\cite{Mano:1996vt}, the asymptotic amplitudes of $R^{\rm in}_{\ell m \omega}$ are built from the coefficient $K_\nu$ [see Eq.(4.2) in Ref.~\cite{Mano:1996vt}], whose numerator carries the factor $\Gamma(1-s-i\varepsilon-i\tau)$ in the notation $\varepsilon=2M\omega$ and $\tau = (\varepsilon-mq)/\sqrt{1-q^2}$ with $q=a/M$.
Using $r_+=M(1+\sqrt{1-q^2})$ and $\kappa=\sqrt{1-q^2}/(2r_+)$ we find
\begin{align}
\begin{split}
    \varepsilon+\tau = \frac{2r_+(\omega-m\Omega_H)}{\sqrt{1-q^2}}=\frac{k_{\rm H}}{\kappa}\,,
\end{split}
\end{align}
so that for $s=-2$ the Gamma function is $\Gamma(3-ik_{\rm H}/\kappa)$, giving rise to the poles at
\begin{align}
\begin{split}
    \omega=m\Omega_{\rm H}-in\kappa,\quad n\ge 3\,.
\end{split}
\end{align}
Both $B^{\rm in}_{\ell m\omega}$ and $B^{\rm out}_{\ell m\omega}$ inherit these simple poles.
$1/B^{\rm in}_{\ell m\omega}$ has zeros there while their ratio, and hence $\hat{D}_{\ell m \omega}$, remains regular.
By Eq.~\eqref{eq:Bin-transform}, the SN inverse amplitude $1/B^{\rm in}_{\rm SN}$ has two more zeros at $n=1$ and $n=2$, supplied by the zeros of $d(\omega)$.
The full set of horizon frequencies with $n\ge1$ therefore appears as zeros of $1/B^{\rm in}_{\rm SN}$, whereas the Teukolsky amplitudes have only those with $n\ge3$.

\bibliography{References}

\end{document}